\documentclass[useAMS,usenatbib]{aa}

\usepackage[varg]{txfonts}
\usepackage{graphicx,color}

\begin{document}

\title{
Chemical (in)homogeneity and atomic diffusion in the open cluster M67\thanks{The data presented herein were obtained at the W.M.\ Keck Observatory, which is operated as a scientific partnership among the California Institute of Technology, the University of California and the National Aeronautics and Space Administration. The Observatory was made possible by the generous financial support of the W.M.\ Keck Foundation.}}

\author{
F. Liu\inst{1}\thanks{E-mail: fan.liu@astro.lu.se}
\and M. Asplund\inst{2,3}
\and D. Yong\inst{2,3}
\and S. Feltzing\inst{1}
\and A. Dotter\inst{4}
\and J. Mel\'endez\inst{5}
\and I. Ram\'irez\inst{6}}

\institute{
Lund Observatory, Department of Astronomy and Theoretical physics, Lund University, Box 43, SE-22100 Lund, Sweden
\and Research School of Astronomy and Astrophysics, Australian National University, Canberra, ACT 2611, Australia
\and ARC Centre of Excellence for All Sky Astrophysics in 3 Dimensions (ASTRO 3D), Australia
\and Harvard-Smithsonian Center for Astrophysics, Cambridge, MA 02138, USA
\and Departamento de Astronomia do IAG/USP, Universidade de Sao Paulo, Rua do Matao 1226, Sao Paulo 05508-900, SP, Brasil
\and Tacoma Community College, Washington, USA}

\date{Received 20 Feb 2019 / Accepted 06 Jun 2019}

\abstract{The benchmark open cluster M67 is known to have solar metallicity and similar age as the Sun. It thus provides us a great opportunity to study the properties of solar twins, as well as the evolution of Sun-like stars.}
{Previous spectroscopic studies reported to detect possible subtle changes in stellar surface abundances throughout the stellar evolutionary phase, namely the effect of atomic diffusion, in M67. In this study we attempt to confirm and quantify more precisely the effect of atomic diffusion, as well as to explore the level of chemical (in)homogeneity in M67.}
{We presented a strictly line-by-line differential chemical abundance analysis of two groups of stars in M67: three turn-off stars and three sub-giants. Stellar atmospheric parameters and elemental abundances were obtained with very high precision using the Keck/HIRES spectra.}
{The sub-giants in our sample show negligible abundance variations ($\le$ 0.02 dex), which implies that M67 was born chemically homogeneous. We note there is a significant abundance difference ($\sim$ 0.1 - 0.2 dex) between sub-giants and turn-off stars, which can be interpreted as the signature of atomic diffusion. Qualitatively stellar models with diffusion agree with the observed abundance results. Some turn-off stars do not follow the general pattern, which suggests that in some cases diffusion can be inhibited, or they might suffered some sort of mixing event related to planets.}
{Our results pose additional challenges for chemical tagging when using turn-off stars. In particular, the effects of atomic diffusion, which could be as large as 0.1 - 0.2 dex, must be taken into account in order for chemical tagging to be successfully applied.}

\keywords{stars: abundances -- stars: atmospheres -- stars: evolution -- Galaxy: open clusters and associations: individual: NGC 2682 (M67)}

\titlerunning{Chemical (in)homogeneity and atomic diffusion in the open cluster M67}
\authorrunning{Liu et al.}
\maketitle

\section{Introduction}

Stars form in clustered environments. It is often assumed that the gas in the proto-clusters is well-mixed, which means that the stars that form out of that gas should all have the same chemical composition (e.g., \citealp{ds06,ds07,fk14}). Open clusters dissolve on time-scales that are short compared to Galactic timescales, thus the clusters are building up the field stellar population when they dissolve. The assumption of chemical homogeneity in the clustered environments where stars form ignites the concept behind chemical tagging. The idea of chemical tagging is that by obtaining high precision elemental abundances we will be able to identify the star-forming events that first provided the stars and/or identify how many such star forming regions have contributed to the stellar field population and help us to recreate, e.g., the star formation history of the stellar disk (\citealp{fb02}). This is a very exciting prospect which merits detailed scrutiny. In a series of investigations we study different cases that will help us to quantify the power of chemical tagging. 

The first aspect to investigate is if open clusters, those that have not yet dissolved, actually are chemically homogenous. We have found that in fact not all clusters are chemically homogenous. We also find that the level of chemical inhomogeneity can be different for light and heavy elements. This was found for the first time in the open cluster Hyades by \citet{liu16a}, with a strictly line-by-line differential analysis and extremely high precision ($\sigma <$ 0.02 dex). It thus appears that if a precision of less than 0.03 dex is needed to chemically tag star forming regions then current surveys might find the task challenging. 

Another aspect that must be accounted for is the fact that many stars have planets and it is now understood that it is possible that planet formation might affect the chemical composition of the host star (for example, see \citealp{mel09,ram10,liu16c}). This could lead to changes in the surface abundance of some elements. \citet{spi18} conducted a high-precision spectroscopic study of five members of the open cluster Pleiades and reported variations in the elemental abundances. They attributed the observed chemical inhomogeneity to the process of forming planets, since the abundance differences correlate with condensation temperature of the elements analysed. One hypothesis is that the refractory materials (elements with high condensation temperature) in the proto-stellar nebula were locked up in the terrestrial planets. The remaining dust-cleansed gas was then accreted on to the host star. Therefore the stars hosting terrestrial planets might be depleted in refractories. Also, planets could be accreted onto their host stars, causing an increase in the abundances of refractory elements.

A third important fact is that even if all stars in an open cluster initially have the same chemical composition, physical processes will change the elemental abundances in their surface layers. One important, and little studied, process that changes the elemental abundances in the surface layers is atomic diffusion. Atomic diffusion is a combination of the effect of gravitational settling, forcing different elements to sink to below the convection zone of the star, and radiative acceleration playing against it \citep{mic84,mic15}. Atomic diffusion is expected to alter the elemental abundances in the upper layers of a star depending on the star's evolutionary stage. The elemental abundances in the surface layers should decrease when the star evolves along the main sequence until they reach the turn-off point. After that the convection zone in the star will deepen and thus will start to restore the original chemical composition in the atmosphere during the sub-giant and red giant phases. Therefore we need to use stellar evolutionary models to infer the original chemical abundances of stars from what we observe \citep{dot17}, in order to use the stars for chemical tagging exercises. 

Several studies have reported on variations of elemental abundances for different evolutionary stages in metal-poor globular clusters like NGC 6397 (see, e.g., \citealp{kor07,lin08,nor12}). Similar phenomenon were detected in NGC 6752 at higher metallicities \citep{gru13}, although with smaller amplitude of the variations in the elemental abundances. However, it is more challenging to test the effect of atomic diffusion in the open clusters since the expected signature is very small particularly for solar mass, solar metallicity, and solar age objects (for example, see \citealp{dot17}). 

M67 is an old benchmark open cluster, which offers great opportunities to study the effect of stellar evolution of Sun-like stars. This metallicity for main sequence stars in M67 have been measured to be similar to that of the Sun \citep{pas08,one11,one14,liu16b}. The age of M67 is about 4 Gyr \citep{yad08,sar09}, also comparable with that of the Sun. This makes M67 an ideal target to probe and test the effect of atomic diffusion and the related stellar evolution models. M67 has been studied by several high resolution spectroscopic studies in recent years. \citet{one11} presented the first high resolution analysis of a solar twin in M67 and found that it has similar chemical composition to that of the Sun. \citet{one14} observed the reduction in surface abundances of heavy elements of the dwarfs and turn-off point stars relative to the sub-giants and they suspected this could be due to the processes of diffusion. \citet{liu16b} studied two solar twins in M67 and confirmed that their chemical abundances are identical to that of the Sun, for the elements with atomic number (Z) $\le$ 30, while one of the solar twin is enriched for the neutron-capture elements, possibly due to the contribution of its binary companion. \citet{sou18} reported the detection of a possible effect of atomic diffusion by studying in total 8 stars in different evolutionary phase, using the APOGEE spectra with resolution of $\sim$ 20,000. Three following studies also found the similar effect using GALAH spectra \citep{gao18}, Gaia-ESO spectra \citep{mot18}, and APOGEE spectra of additional stars \citep{sou19}. In these studies, the abundance differences between the turn-off and sub-giant stars were 0.05 - 0.07 dex, comparable to the typical measurement uncertainties. A more compelling demonstration of atomic diffusion in this cluster would be provided if the abundance errors were a factor of 2 - 3 smaller than the atomic diffusion signature. 

In this paper, we presented a strictly line-by-line differential elemental abundance analysis of three turn-off stars and three sub-giant stars in the old open cluster M67. For the analysis we have obtained high resolution, very high signal-to-noise ratio (SNR) spectra. We use these spectra to examine the possible effect of atomic diffusion in this old open cluster.

\section{Observations and data reduction}

\begin{figure}
\centering
\includegraphics[width=\columnwidth]{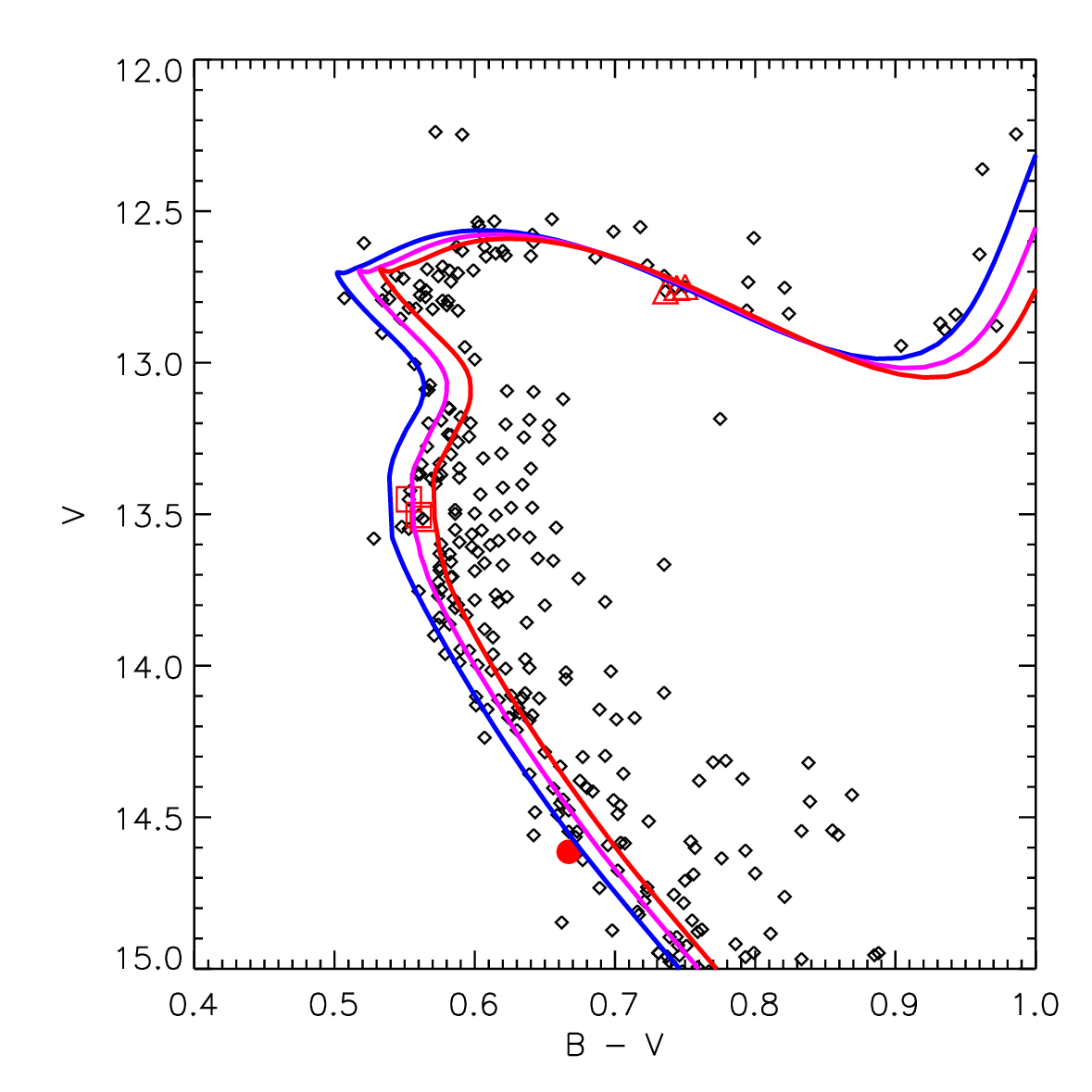}
\caption{Our programme stars in the colour magnitude diagram, which were selected from \citet{yad08}. They are all high probability members. The red symbols correspond to our targets: solar twin Y1194 (red circle), turn-off stars (red rectangles), sub-giant stars (red triangles), respectively. The blue, magenta, and red solid lines correspond to the isochrones for an age of 4 Gyr, $(m-M)_0$ = 9.70, and initial metallicities of [Fe/H] = 0.0, 0.05, and 0.1 from MIST \citep{dot16,cho16}, respectively.}
\label{fig1}
\end{figure}

We carefully selected our targets depending on their position in the colour-magnitude diagram (see Figure \ref{fig1}). In particular we selected stars in two groups: turn-off stars and sub-giants. We note that our programme stars, taken from \citet{yad08}, are all high probability members of M67 based on their proper motions. According to the Gaia Data Release 2 \citep{gai18}, our sample stars also have very similar parallaxes, around 1.0848\,$mas$, further confirming their membership. Basic data for our sample stars are listed in Table \ref{t:info}.

\begin{table*}
\caption{Information of our programme stars.}
\centering
\label{t:info}
\begin{tabular}{@{}lccccccc@{}}
\hline\hline
Object & 2MASS ID & $B$ & $V$ & Proper motions$^a$ & Parallaxes$^a$ & Probability$^b$& Type of\\
 & & mag & mag & $mas$\,yr$^{-1}$ & $mas$ & \% & star$^c$\\
\hline
Y1194 & J08510080+1148527 & 15.28 & 14.61 & $-$10.878 $-$2.757 [0.067 0.061] & 1.1164 [0.0369] & 99 & ST\\
Y519 & J08510951+1141449 & 13.50 & 12.75 & $-$10.333 $-$3.106 [0.075 0.059] & 1.0848 [0.0431] & 98 & SG \\
Y923 & J08514401+1146245 & 13.50 & 12.75 & $-$11.100 $-$2.889 [0.072 0.046] & 1.1190 [0.0415] & 100 & SG \\
Y1844 & J08513540+1157564 & 13.50 & 12.76 & $-$11.097 $-$3.008 [0.067 0.044] & 1.1499 [0.0422] & 91 & SG\\
Y535 & J08510131+1141587 & 14.00 & 13.45 & $-$11.041 $-$2.813 [0.045 0.031] & 1.1077 [0.0261] & 100 & TO\\
Y1388 & J08505474+1151093 & 14.08 & 13.52 & $-$10.821 $-$2.941 [0.046 0.036] & 1.0874 [0.0273] & 99 & TO\\
Y2235 & J08510470+1204193 & 14.06 & 13.50 & $-$10.944 $-$2.967 [0.041 0.029] & 1.1506 [0.0262] & 100 &TO \\
\hline
\end{tabular}
\\
$^a$ Proper motions (in RA and Dec), and parallaxes with uncertainties were taken from \citet{gai18}.\\
$^b$ Probability of being a member of M67, taken from \citet{yad08}.\\
$^c$ Stellar type: sub-giant star (SG), turn-off star (TO), solar twin (ST).
\end{table*}

For this project, we observed three turn-off stars (Y1388, Y535, and Y2235) and three sub-giant stars (Y1844, Y519, and Y923). We obtained the high resolution (R = $\lambda/\Delta\lambda$ = 50,000), high SNR spectra with the 0.86" slit, \textit{kv408} filter of the High Resolution Echelle Spectrometer (HIRES, \citealp{vog94}) on the 10 m Keck I telescope during the nights of February 1, April 8 and April 9, 2017. The spectral wavelength coverage is nearly complete from 400 to 840 nm.

\begin{figure}
\centering
\includegraphics[width=\columnwidth]{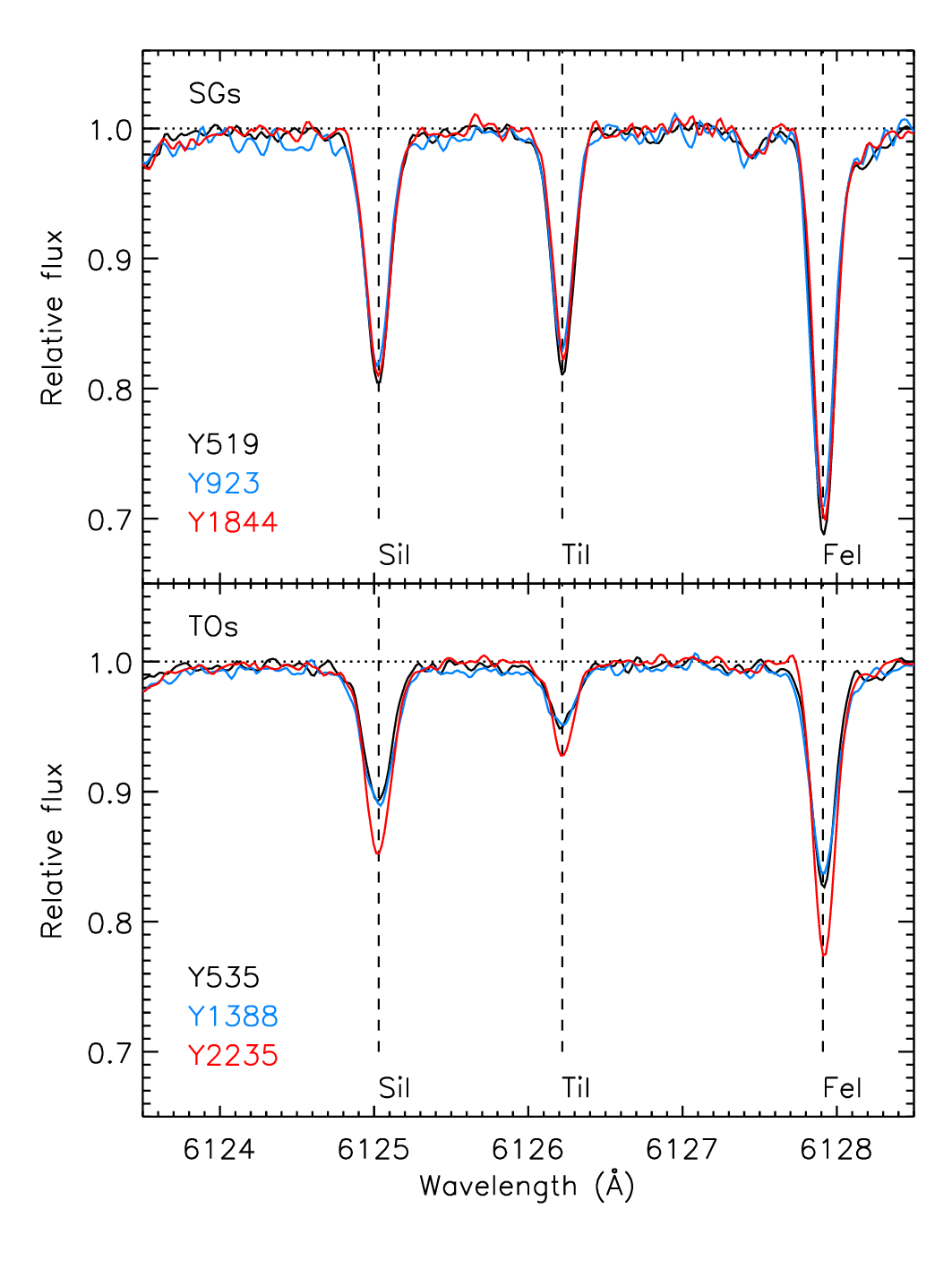}
\caption{Top panel: Portion of the reduced spectra for the sub-giant stars in our sample; Y519, Y923, and Y1844 are plotted in black, blue, and red, respectively. A few atomic lines (Si\,{\sc i}, Ti\,{\sc i}, Fe\,{\sc i}) used in our analysis in this region are marked by the dashed lines. Bottom panel: similar to the top panel, but for the turn-off stars in our sample; Y535, Y1388, and Y2235 are plotted in black, blue, and red, respectively.}
\label{fig2}
\end{figure}

The Keck-MAKEE pipeline was used for standard echelle spectra reduction including bias subtraction, flat-fielding, scattered-light subtraction, spectra extraction and wavelength calibration. We normalized, barycentric velocity corrected and co-added the spectra using packages within IRAF\footnote{IRAF is distributed by the National Optical Astronomy Observatory, which is operated by Association of Universities for Research in Astronomy, Inc., under cooperative agreement with National Science Foundation.}. The spectrum of each star has been radial velocity (RV) calibrated to the rest wavelength, using the \textit{rv} package of IRAF. The individual frames of each star were combined into a single spectrum with SNR $\approx$ 250 - 300 per pixel in most wavelength regions. We note that one of our programme star (Y535) shows RV variation, when compared to the other stars, which indicates that this star might has a wide binary companion. We note that no double-line features could be identified in the spectrum of Y535. Portions of the reduced spectra for the sub-giant and turn-off stars are shown in Figure \ref{fig2}.

We also included a solar spectrum by observing the asteroid Hebe (SNR $\approx$ 450 per pixel), as well as a spectrum of main sequence solar twin in M67 (Y1194, SNR $\approx$ 270 per pixel) in this study. These two spectra\footnote{These two spectra were used in our previous study \citep{liu16b}, but re-analysed in this study for comparison and consistency purposes.} share the same instrumental configuration as the spectra of our program stars, and reduced using the same method. 

The spectral line-list of 22 elements (C, O, Na, Mg, Al, Si, S, Ca, Sc, Ti, V, Cr, Mn, Fe, Co, Ni, Cu, Zn, Sr, Y, Ba, and Ce) employed in our analysis was adopted mainly from \citet{sco15a,sco15b} and \citet{gre15}, and complemented with additional unblended lines from \citet{ben05} and \citet{mel14}. Equivalent widths were measured manually using the \textit{splot} task in IRAF. We note that each spectral line was measured consecutively for all the programme stars by setting a consistent continuum, resulting in precise measurements in a differential sense. Strong lines with EW $\geq$ 120 m\AA\ were excluded from the analysis to limit the effects of saturation with the exception of a few Mg\,{\sc i}, Mn\,{\sc i,} and Ba\,{\sc i} lines. The atomic line data as well as the measured equivalent widths we adopted for our analysis are listed in Table A1.

\section{Analysis and results}

\subsection{Stellar atmospheric parameters}

\begin{figure}
\centering
\includegraphics[width=\columnwidth]{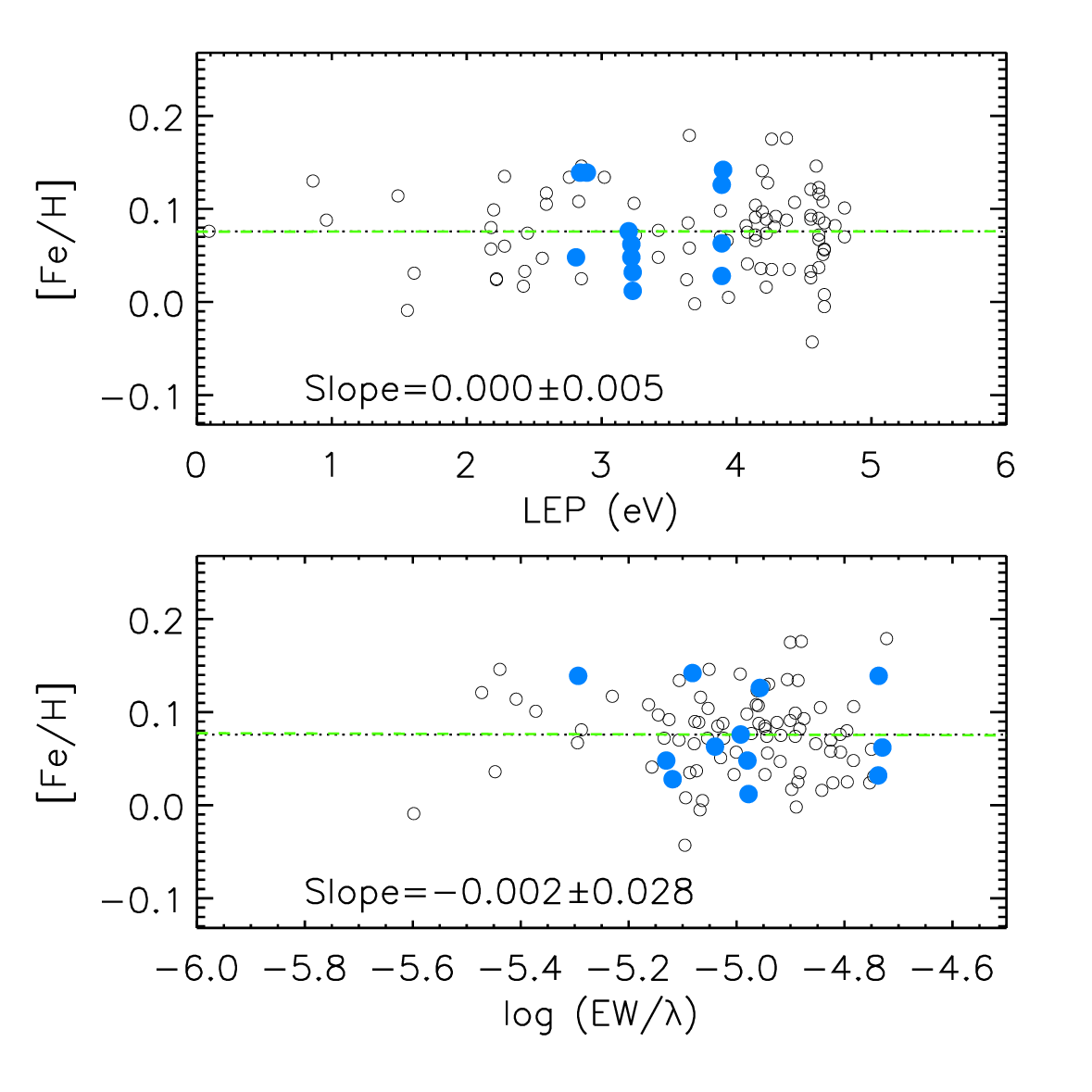}
\caption{Top panel: [Fe/H] of Y923 derived on a line-by-line basis with respect to the Sun as a function of LEP; open circles and blue filled circles represent Fe\,{\sc i} and Fe\,{\sc ii} lines, respectively. The black dotted line shows the location of mean [Fe/H], the green dashed line represents the linear least-squares fit to the data. Bottom panel: same as in the top panel but as a function of reduced equivalent width.}
\label{fig3}
\end{figure}

We conducted a 1D, local thermodynamic equilibrium (LTE) elemental abundance analysis using MOOG (version 2014, \citealp{sne73,sob11}) with the Kurucz model atmospheres (ODFNEW grid, \citealp{cas03}). Stellar atmospheric parameters (i.e., effective temperature T$_{\rm eff}$, surface gravity $\log g$, microturbulent velocity $\xi_{\rm t}$, and metallicity [Fe/H]) were obtained by forcing differential excitation and ionization balance of Fe\,{\sc i} and Fe\,{\sc ii} lines on a strictly line-by-line basis relative to the Sun (reflected light from the Hebe asteroid). The adopted parameters for the Sun are fixed as: T$_{\rm eff} = 5772$\,K, $\log g$ = 4.44 [cm\,s$^{-2}$], $\xi_{\rm t}$ = 1.00 km\,s$^{-1}$, and [Fe/H] = 0.00. The stellar parameters of our programme stars were derived individually using an automatic grid searching technique described in \citet{liu14}. Briefly, the best combination of stellar atmospheric parameters, minimizing the slopes in [Fe\,{\sc i}/H] versus lower excitation potential (LEP) and reduced equivalent width ($\log$\,(EW/$\lambda$) as well as the difference between [Fe\,{\sc i}/H] and [Fe\,{\sc ii}/H], was determined from a successively refined grid of stellar atmospheric models. The final solution was achieved when the grid step-size decreased to $\Delta$T$_{\rm eff}$ = 1 K, $\Delta \log g$ = 0.01 [cm\,s$^{-2}$] and $\Delta$$\xi_{\rm t}$ = 0.01 km\,s$^{-1}$. We also require the derived average [Fe/H] to be consistent with the adopted model atmospheric value. Lines whose elemental abundances departed from the average by $> 3\,\sigma$ were clipped. Figure \ref{fig3} shows an example of obtaining the stellar atmospheric parameters of the sub-giant star Y923 relative to the Sun. The adopted stellar parameters satisfy the excitation and ionization balance in a differential sense.

The final adopted stellar atmospheric parameters of our program stars, as well as the data for the solar twin Y1194 and the Sun and their parameters as a reference point, are listed in Table \ref{t:para}. The adopted uncertainties in the stellar parameters were calculated using the method described by \citet{eps10} and \citet{ben14}, which accounts for the co-variances between changes in the stellar parameters and the differential iron abundances. High precision was achieved due to the high quality spectra and the strictly line-by-line differential method, which greatly reduces the systematic errors from atomic line data, shortcomings in the 1D LTE modelling of the stellar atmospheres and spectral line formation (see, e.g., \citealp{asp09}). In this study, we are able to reproduce the results for the solar twin Y1194 as in our previous study \citep{liu16b}. In addition, we repeated the whole analysis by choosing a typical sub-giant star (Y923) and a typical turn-off star (Y1388) as a reference, respectively. The results are essentially the same with slightly different uncertainties and do not change our final results and conclusion. Since the parameters of the Sun are closer to both turn-off and sub-giant stars, leading to the results with slightly smaller uncertainties, we choose the Sun as the reference star in the following analyses. We note that the overall metallicities of the sub-giant stars are higher than that of the turn-off stars by more than 0.1 dex. We Also note that the turn-off star Y2235 has a higher metallicity, when compared to the other two other turn-off stars. Detailed discussion on this "outlier" is presented in the Section\,\ref{sect:abunTO}.

We compared the stellar parameters derived from this study to several previous spectroscopic studies for the stars in common. Our turn-off star Y1388 has been studied in \citet{gao18} and our sub-giant star Y1844 in the two studies \citep{gao18,sou18}. For both stars, our results agree well with the previous studies in terms of T$_{\rm eff}$ and $\log g$ with consideration of the estimated errors. The metallicity ([Fe/H]) for Y1388 in this study is higher than that from \citet{gao18} by $\approx$ 0.07 dex, while [Fe/H] for Y1844 in this study is higher by $\approx$ 0.08 dex and 0.05 dex than that from \citet{gao18} and \citet{sou18}, respectively. The differences in [Fe/H] are most likely due to a zero-point offset, which has essentially little impact to the conclusion. As mentioned in the Introduction, we have achieved our goal of obtaining precision 2 - 3 times better than the previous studies, mainly because our spectra have higher resolution (especially when compared to APOGEE spectra), as well as higher signal-to-noise ratios ($250 - 300$ in this study; $50 - 150$ in previous studies). 

\begin{table*}
\caption{Adopted solar parameters and derived stellar parameters for our programme stars (relative to the Sun). The last column gives the type of star: sub-giant star (SG), turn-off star (TO), and solar twin on the main-sequence (ST).}
\centering
\label{t:para}
\begin{tabular}{@{}lccccc@{}}
\hline\hline
Object & T$_{\rm eff}$ & $\log g$ & $\xi_{\rm t}$ & [Fe/H] & \\
 & (K) & [cm\,s$^{-2}$] & (km\,s$^{-1}$) & & \\
\hline
Hebe & 5772 & 4.44 & 1.00 & 0.00 & Sun \\
\hline
Y1194 & 5744 $\pm$ 11 & 4.43 $\pm$ 0.03 & 0.98 $\pm$ 0.03 & $-$0.005 $\pm$ 0.012 & ST \\
Y519 & 5586 $\pm$ 18 & 3.83 $\pm$ 0.05 & 1.23 $\pm$ 0.03 & 0.084 $\pm$ 0.016 & SG \\
Y923 & 5651 $\pm$ 18 & 3.88 $\pm$ 0.06 & 1.27 $\pm$ 0.04 & 0.076 $\pm$ 0.019 & SG \\
Y1844 & 5628 $\pm$ 19 & 3.84 $\pm$ 0.06 & 1.26 $\pm$ 0.04 & 0.072 $\pm$ 0.020 & SG \\
Y535 & 6155 $\pm$ 23 & 4.18 $\pm$ 0.05 & 1.42 $\pm$ 0.06 & $-$0.128 $\pm$ 0.017 & TO \\
Y1388 & 6124 $\pm$ 25 & 4.12 $\pm$ 0.05 & 1.36 $\pm$ 0.05 & $-$0.079 $\pm$ 0.020 & TO \\
Y2235 & 6098 $\pm$ 21 & 4.21 $\pm$ 0.05 & 1.34 $\pm$ 0.04 & 0.041 $\pm$ 0.018 & TO \\
\hline
\end{tabular}
\end{table*}

\subsection{Elemental abundances}

Having established the stellar atmospheric parameters of our programme stars, we then derived abundances for 21 elements in addition to Fe (C, O, Na, Mg, Al, Si, S, Ca, Sc, Ti, V, Cr, Mn, Co, Ni, Cu, Zn, Sr, Y, Ba, and Ce) based on the spectral lines and measured equivalent widths listed in Table A1. We derived line-by-line differential elemental abundances ([X/H]) of our programme stars relative to the Sun. We note that an element with different ionization stages was considered as two species in our analysis, rather than combining their elemental abundances together. We took hyperfine structure (HFS) into account for 6 elements (Sc, V, Mn, Co, Cu, and Ba) and calculated corrections strictly line-by-line. The HFS data were taken from \citet{kur95} for Sc, V, Co, Cu and Ba. For Mn, the HFS corrections were applied using the data from \citet{pro00} and \citet{bb15}. The average corrections are smaller than 0.01 dex for Sc, V, Cu, and Ba. However for Mn, the average corrections are $\approx$ $-$0.06 dex for sub-giant stars and $\approx$ $+$0.09 dex for turn-off stars; while for Co, the average corrections are $\approx$ $-$0.02 dex for the sub-giant stars and $\approx$ $+$0.03 dex for turn-off stars.

We then applied differential non-LTE (NLTE) corrections\footnote{All the corrections were made line-by-line differentially relative to the Sun.} for our sample stars for those essential elements (O, Na, Mg, Al, Si, Ti and Fe) for this study. Below, we briefly describe the process used for each element. 

\textit{Oxygen.} We adopted 3D NLTE corrections for the abundance determination using the 777nm triplet. The corrections were based on \citet{ama16}. The differential 3D NLTE abundance corrections for oxygen are about $-$0.100 dex for the sub-giant stars and $-$0.134 dex for the turn-off stars for these lines. These corrections are substantial and should not be ignored. 

\textit{Sodium.} For the Na lines at 615.4nm and 616.0nm 1D NLTE corrections were calculated using the INSPECT database\footnote{www.inspect-stars.com}. These corrections are based on \citet{lin11}. The differential 1D NLTE abundance corrections for Na are about $-$0.009 dex for the sub-giant stars and $-$0.008 dex for the turn-off stars. 

\textit{Magnesium.} For the Mg lines at 571.1nm, 631.8nm and 631.9nm the differential 1D NLTE abundance corrections were calculated using the GUI web-tool from Maria Bergemann's group\footnote{http://nlte.mpia.de/gui-siuAC\_sec.php}. These corrections are based on \citet{ber15}. The differential 1D NLTE abundance corrections for Mg are almost zero for the sub-giant stars and about $+$0.013 dex for the turn-off stars. 

\textit{Aluminium.} For the Al lines at 669.6nm and 669.8nm we adopted $<$3D$>$ NLTE corrections from \citet{nl17}. The differential $<$3D$>$ NLTE abundance correction for Al are about $-$0.013 dex for the sub-giant stars and $+$0.017 dex for the turn-off stars. 

\textit{Silicon, titannium, and iron.} 1D NLTE corrections for our Si, Ti\,{\sc i}, and Fe\,{\sc i} lines were derived using the GUI web-tool based on \citet{ber13}, \citet{ber11}, and \citet{ber12}, respectively. The differential 1D NLTE abundance corrections for Si are about $-$0.005 dex for the sub-giant stars and $-$0.003 dex for the turn-off stars. For Ti\,{\sc i}, the abundance corrections are almost zero for the sub-giant stars and $+$0.030 dex for the turn-off stars. For Fe\,{\sc i}, the abundance corrections are $+$0.002 dex for the sub-giant stars and $+$0.005 dex for the turn-off stars. 

\begin{figure}
\centering
\includegraphics[width=0.95\columnwidth]{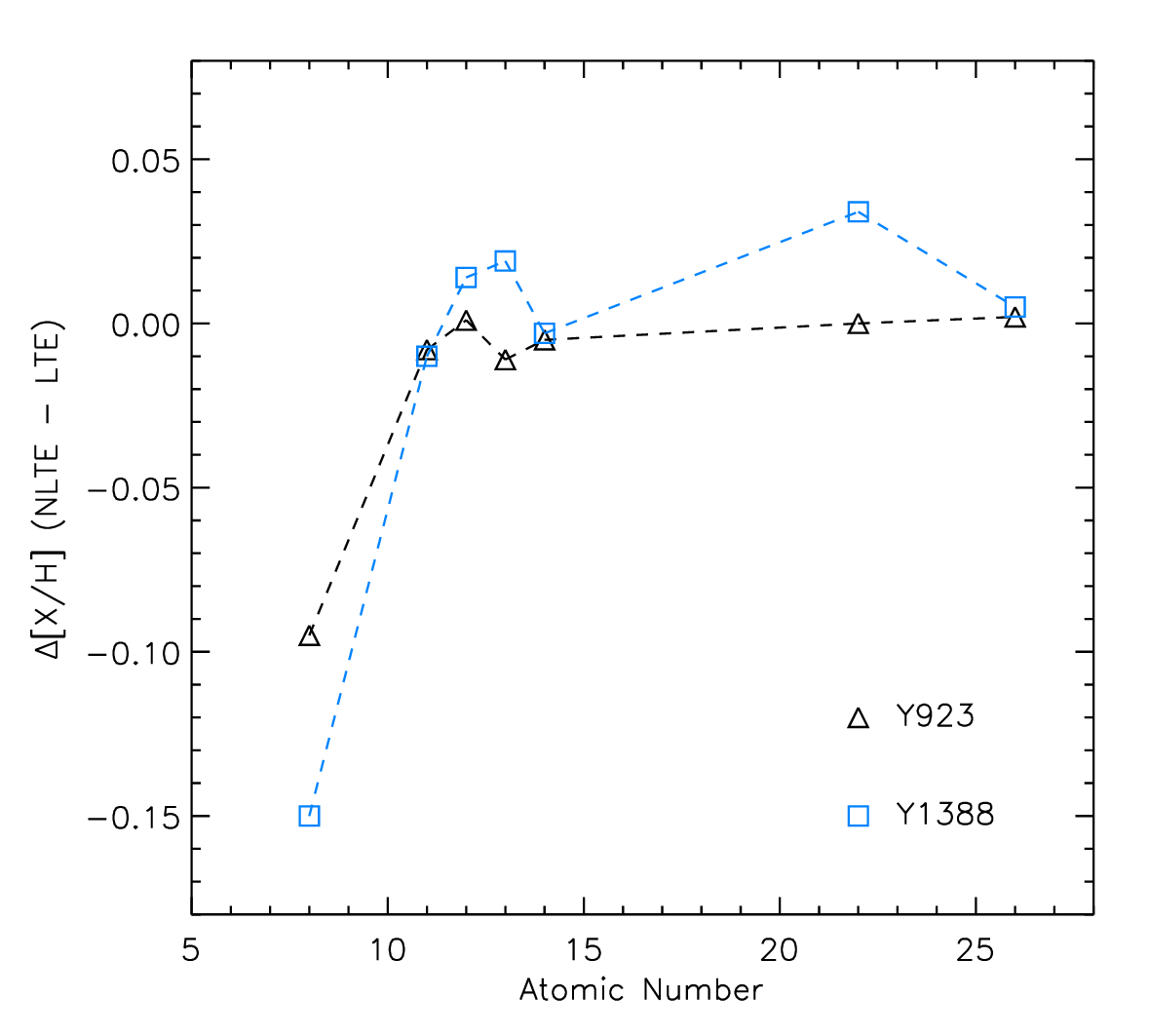}
\caption{NLTE corrections $\Delta$[X/H] (NLTE $-$ LTE) for the sub-giant star Y923 (black triangles) and the turn-off star Y1388 (blue rectangles), as a function of atomic number. The corrections were calculated line-by-line relative to the Sun.}
\label{fig4}
\end{figure}

Figure \ref{fig4} shows the NLTE corrections for the sub-giant star Y923 and the turn-off star Y1388 relative to the Sun, as a function of atomic number. We can tell that in a differential sense, only the NLTE corrections for the oxygen triplet are significant enough to alter our final results. The amplitude of the NLTE corrections in our study agree with that from \citet{gao18}, except for Na, where they found large difference (0.18 dex) between NLTE and LTE abundances. However their corrections are based on two different Na lines.

Table \ref{t:abun} lists the adopted differential chemical abundances of all our sample stars relative to the Sun for in total 22 elements (24 species). The errors in the elemental abundances were calculated following the method described by \citet{eps10}: the standard errors in the mean abundances, as derived from the different spectral lines, were added in quadrature to the errors introduced by the uncertainties in the stellar atmospheric parameters. For elements where only one spectral line was measured (i.e., S, Sr and Ce), we estimate the uncertainties by taking into consideration errors due to the signal-to-noise ratio, the continuum setting and the stellar parameters. The quadratic sum of the three uncertainties sources give the errors for these two elements. The inferred errors on differential elemental abundances are listed in Table \ref{t:abun}. Most derived elemental abundances have uncertainties $\le$ 0.025 dex, which further underscores the advantages of a strictly differential abundance analysis. When considering all species, the average uncertainties are 0.023 $\pm$ 0.002 ($\sigma$ = 0.008) dex for the sub-giant stars and 0.025 $\pm$ 0.002 ($\sigma$ = 0.009) dex for the turn-off stars, relative to the Sun. We note that the errors in stellar parameters and elemental abundances for the sub-giant stars are slightly smaller than that for the turn-off stars.

\begin{table*}
\caption{Differential chemical abundances of our sample stars for 22 elements (24 species), relative to the Sun.}
\centering
\label{t:abun}
\begin{tabular}{@{}cccc|ccc@{}}
\hline
 & \multicolumn{3}{c|}{[X/H] (sub-giant stars)} & \multicolumn{3}{c}{[X/H] (turn-off stars)} \\
Species & Y519 & Y923 & Y1844 & Y535 & Y1388 & Y2235 \\
\hline
C\,{\sc i} & 0.091 $\pm$ 0.022 & 0.014 $\pm$ 0.018 &0.041 $\pm$ 0.012 &$-$0.111 $\pm$ 0.024 &$-$0.063 $\pm$ 0.023 & 0.013 $\pm$ 0.021\\
O\,{\sc i}$^a$ & 0.048 $\pm$ 0.025 & 0.023 $\pm$ 0.024 &0.068 $\pm$ 0.027 &$-$0.096 $\pm$ 0.023 &$-$0.047 $\pm$ 0.027 & 0.023 $\pm$ 0.019\\
Na\,{\sc i}$^a$ & 0.080 $\pm$ 0.018 & 0.063 $\pm$ 0.021 &0.054 $\pm$ 0.009 &$-$0.122 $\pm$ 0.013 &$-$0.084 $\pm$ 0.016 & 0.021 $\pm$ 0.008\\
Mg\,{\sc i}$^a$ & 0.132 $\pm$ 0.010 & 0.103 $\pm$ 0.006 &0.123 $\pm$ 0.008 &$-$0.117 $\pm$ 0.015 &$-$0.067 $\pm$ 0.009 & 0.066 $\pm$ 0.012\\
Al\,{\sc i}$^a$ & 0.091 $\pm$ 0.008 & 0.053 $\pm$ 0.007 &0.085 $\pm$ 0.015 &$-$0.179 $\pm$ 0.008 &$-$0.128 $\pm$ 0.038 &$-$0.014 $\pm$ 0.011\\
Si\,{\sc i}$^a$ & 0.117 $\pm$ 0.008 & 0.099 $\pm$ 0.014 &0.088 $\pm$ 0.010 &$-$0.099 $\pm$ 0.011 &$-$0.052 $\pm$ 0.009 & 0.047 $\pm$ 0.009\\
 S\,{\sc i} & 0.102 $\pm$ 0.026 & 0.082 $\pm$ 0.027 &0.073 $\pm$ 0.027 &$-$0.100 $\pm$ 0.026 &$-$0.084 $\pm$ 0.027 &$-$0.030 $\pm$ 0.026\\
Ca\,{\sc i} & 0.093 $\pm$ 0.014 & 0.074 $\pm$ 0.014 &0.082 $\pm$ 0.014 &$-$0.087 $\pm$ 0.017 &$-$0.056 $\pm$ 0.016 & 0.052 $\pm$ 0.017\\
Sc\,{\sc ii}& 0.096 $\pm$ 0.022 & 0.085 $\pm$ 0.026 &0.102 $\pm$ 0.028 &$-$0.114 $\pm$ 0.023 &$-$0.105 $\pm$ 0.028 & 0.045 $\pm$ 0.026\\
Ti\,{\sc i}$^a$ & 0.058 $\pm$ 0.022 & 0.062 $\pm$ 0.024 &0.048 $\pm$ 0.022 &$-$0.100 $\pm$ 0.021 &$-$0.085 $\pm$ 0.023 & 0.015 $\pm$ 0.020\\
Ti\,{\sc ii}& 0.105 $\pm$ 0.021 & 0.083 $\pm$ 0.024 &0.082 $\pm$ 0.023 &$-$0.125 $\pm$ 0.032 &$-$0.127 $\pm$ 0.026 & 0.023 $\pm$ 0.028\\
 V\,{\sc i} & 0.083 $\pm$ 0.020 & 0.127 $\pm$ 0.022 &0.086 $\pm$ 0.032 &$-$0.088 $\pm$ 0.024 &$-$0.079 $\pm$ 0.029 & 0.031 $\pm$ 0.032\\
Cr\,{\sc i} & 0.074 $\pm$ 0.015 & 0.062 $\pm$ 0.023 &0.047 $\pm$ 0.019 &$-$0.122 $\pm$ 0.024 &$-$0.081 $\pm$ 0.026 & 0.003 $\pm$ 0.021\\
Mn\,{\sc i} & 0.065 $\pm$ 0.022 & 0.012 $\pm$ 0.021 &0.015 $\pm$ 0.029 &$-$0.170 $\pm$ 0.027 &$-$0.158 $\pm$ 0.032 &$-$0.011 $\pm$ 0.025\\
Fe\,{\sc i}$^a$ & 0.085 $\pm$ 0.012 & 0.078 $\pm$ 0.013 &0.074 $\pm$ 0.013 &$-$0.122 $\pm$ 0.015 &$-$0.074 $\pm$ 0.016 & 0.046 $\pm$ 0.015\\
Fe\,{\sc ii}& 0.085 $\pm$ 0.018 & 0.076 $\pm$ 0.023 &0.074 $\pm$ 0.023 &$-$0.127 $\pm$ 0.019 &$-$0.079 $\pm$ 0.023 & 0.041 $\pm$ 0.020\\
Co\,{\sc i} & 0.091 $\pm$ 0.020 & 0.064 $\pm$ 0.022 &0.032 $\pm$ 0.024 &$-$0.077 $\pm$ 0.017 &$-$0.083 $\pm$ 0.018 & 0.011 $\pm$ 0.022\\
Ni\,{\sc i} & 0.080 $\pm$ 0.013 & 0.078 $\pm$ 0.015 &0.072 $\pm$ 0.014 &$-$0.145 $\pm$ 0.017 &$-$0.099 $\pm$ 0.018 & 0.023 $\pm$ 0.018\\
Cu\,{\sc i} & 0.113 $\pm$ 0.039 & 0.065 $\pm$ 0.039 &0.085 $\pm$ 0.026 &$-$0.173 $\pm$ 0.022 &$-$0.128 $\pm$ 0.040 & 0.040 $\pm$ 0.023\\
Zn\,{\sc i} &$-$0.004 $\pm$ 0.030 &$-$0.010 $\pm$ 0.030 &0.034 $\pm$ 0.013 &$-$0.210 $\pm$ 0.031 &$-$0.137 $\pm$ 0.019 &$-$0.053 $\pm$ 0.022\\
Sr\,{\sc i} & 0.039 $\pm$ 0.031 & 0.051 $\pm$ 0.032 &0.067 $\pm$ 0.032 &$-$0.074 $\pm$ 0.031 &$-$0.071 $\pm$ 0.032 & 0.021 $\pm$ 0.032\\
 Y\,{\sc ii}& 0.095 $\pm$ 0.023 & 0.058 $\pm$ 0.031 &0.087 $\pm$ 0.029 &$-$0.145 $\pm$ 0.024 &$-$0.124 $\pm$ 0.026 & 0.030 $\pm$ 0.025\\
Ba\,{\sc ii}& 0.117 $\pm$ 0.024 & 0.111 $\pm$ 0.030 &0.100 $\pm$ 0.025 &$-$0.128 $\pm$ 0.034 &$-$0.063 $\pm$ 0.039 & 0.069 $\pm$ 0.030\\
Ce\,{\sc ii}& 0.086 $\pm$ 0.033 & 0.079 $\pm$ 0.037 &0.077 $\pm$ 0.037 &$-$0.129 $\pm$ 0.035 &$-$0.134 $\pm$ 0.036 &$-$0.022 $\pm$ 0.034\\
\hline
\end{tabular}
\\
$^a$ Abundances are NLTE corrected, line-by-line relative to the Sun.
\end{table*}

\section{Discussion}

\subsection{Elemental abundances of the sub-giant stars in M67}

\begin{figure}
\centering
\includegraphics[width=\columnwidth]{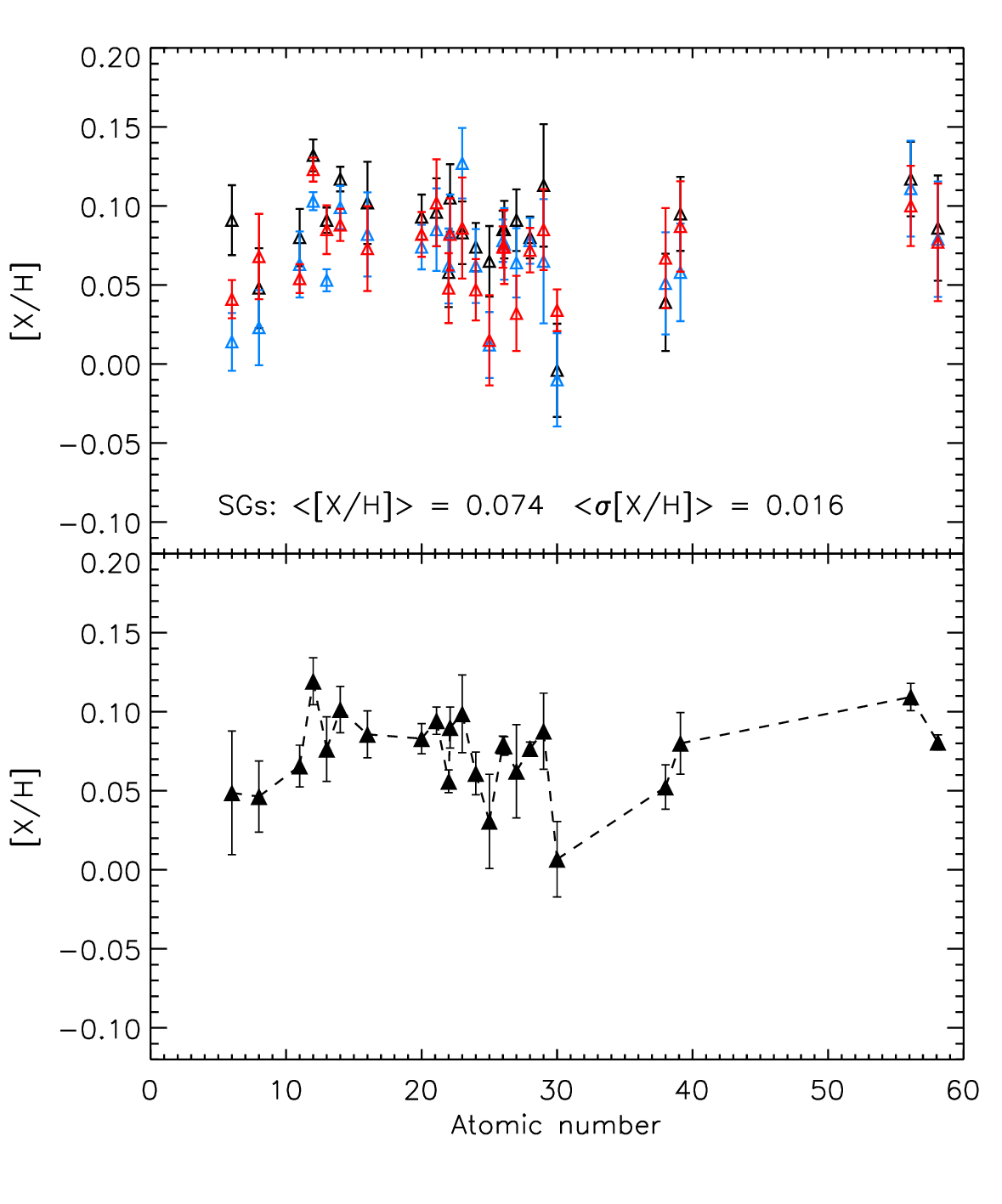}
\caption{Top panel: Elemental abundances of our sub-giant stars as a function of atomic number; black, blue, and red triangles represent the elemental abundances of Y519, Y923 and Y1844, respectively. Bottom panel: Similar to the top panel but for the average abundances of the sub-giant stars (black filled triangles). The error bars represent the standard deviations of the mean values.}
\label{fig5}
\end{figure}

It is of particular interest to explore the elemental abundances of all our sample stars, as well as the abundance variations within each stellar group. Figure \ref{fig5} shows the elemental abundances of our sub-giant stars (relative to the Sun) as a function of atomic number. In order to clarify the quantities of chemical homogeneity in the sub-giant stars, we list the average elemental abundances, related dispersions (standard deviations of the mean), and the corresponding average errors for the sub-giant stars in our sample for the 24 species in Table \ref{t:stat}. We note that the dispersions are comparable or smaller than the average errors for almost all the elements. The average elemental abundance of all the elements for the three sub-giant stars in our sample is 0.074 dex, with an average dispersion of 0.016 dex, and an average error of 0.022 dex. Therefore, in spite of our small sample size, we would argue that the sub-giant stars in our sample are chemically homogenous, also at a extremely high precision level ($\sim$ 0.02 dex).

\subsection{Elemental abundances of turn-off stars in M67}
\label{sect:abunTO}

\begin{figure}
\centering
\includegraphics[width=\columnwidth]{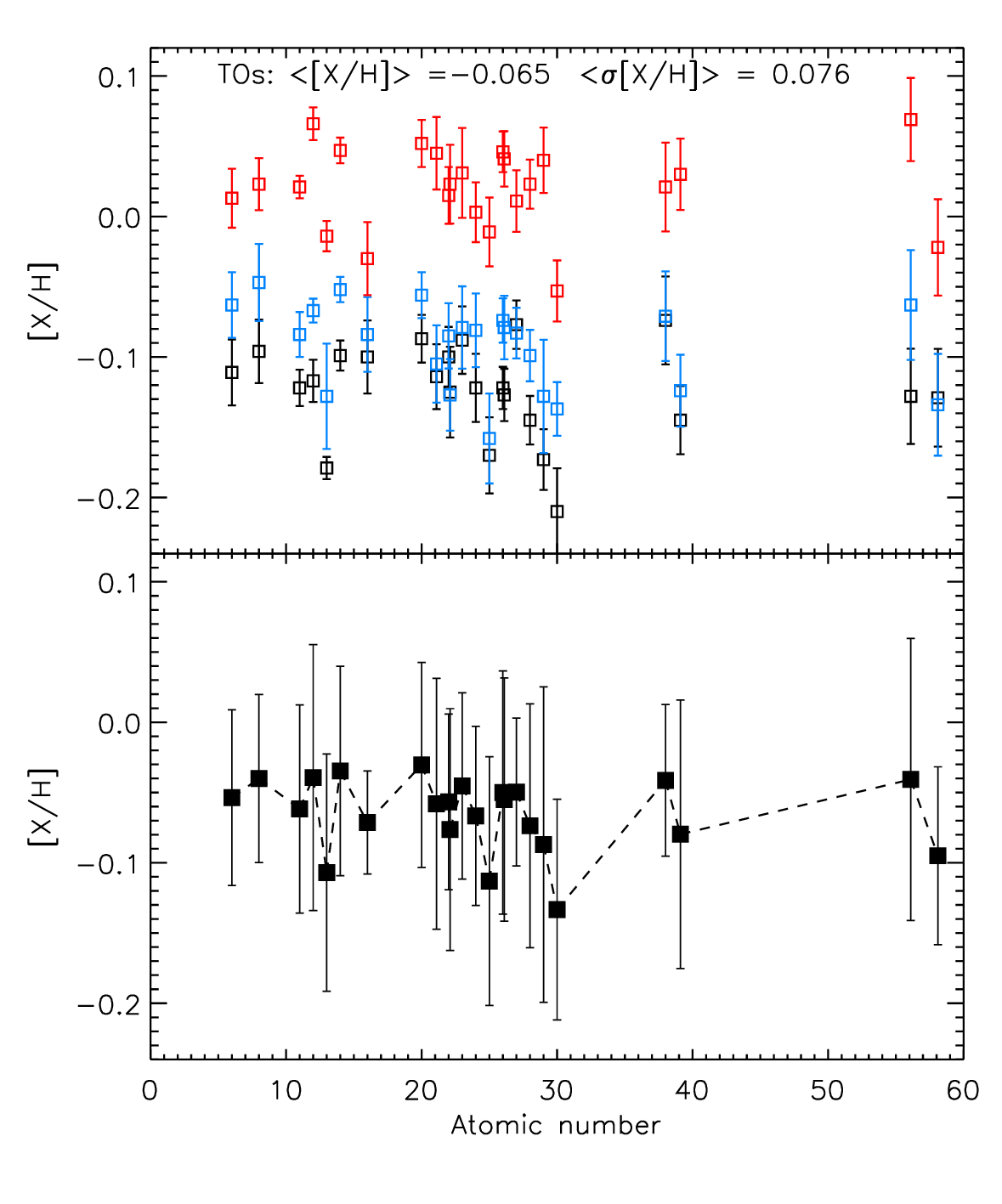}
\caption{Top panel: Elemental abundances of the turn-off stars in our sample as a function of atomic number: black, blue, and red rectangles represent the abundances of Y535, Y1388, and Y2235, respectively. Bottom panel: Similar to the top panel but for the average elemental abundances of the turn-off stars (black filled rectangles). The error bars represent the standard deviations of the mean values.}
\label{fig6}
\end{figure}

The elemental abundances of turn-off stars (relative to the Sun) in our sample as a function of atomic number is shown in Figure \ref{fig6}. The average elemental abundances, related dispersions and the average errors for the turn-off stars in our sample are listed in Table\,\ref{t:stat}. It is clear that the variations in the elemental abundances in our turn-off stars are much larger than those of the sub-giant stars. For the turn-off stars we find an average elemental abundance of $-$0.065 dex and an average dispersion in of 0.076 dex, while the average error is 0.023 dex. According to Table \ref{t:stat}, all the species have elemental abundance dispersions 2 - 4 times larger than the average errors.

At a first glance, such a significant variation in elemental abundances in turn-off stars might be due to one "outlier", namely Y2235. Y2235 has an overall metallicity of $\approx$ 0.04 dex, about 0.12 dex higher than that of Y1388 and 0.17 dex higher than that of Y535. However, we did not find any indications of binarity, i.e. RV variation or visible double line feature from the spectrum of this star, that could help explain the odd elemental abundances. Y2235 has a slightly cooler T$_{\rm eff}$ and a slightly larger $\log g$ when compared to the other two turn-off stars. It is not likely that such a small difference in stellar parameters can cause such a large the change in elemental abundance ($>$ 0.1 dex). We would have to decrease the T$_{\rm eff}$ by about 150 K (7 times than the uncertainty in T$_{\rm eff}$) or decrease the $\log g$ by about 0.2 dex (4 times than the uncertainty in $\log g$), in order to force the metallicity of Y2235 to drop by about 0.1 dex, ignoring the impact of those changes on the excitation/ionization balance essential for obtaining the stellar parameters. Such a significant change would be neither realistic nor reasonable. In addition, by checking the spectrum, we note that Y2235 does have deeper spectral lines when compared to the two other turn-off stars (see Figure \ref{fig2}). We can not exclude the possibility that Y2235 has suffered some sort of mixing event related to planet engulfment. Since this is not the main scope of this paper, we only discuss briefly the planet related scenarios in Appendix B, in order to avoid the distraction of the readers.

We also noticed that the most metal-poor turn-off star, Y535, has a metallicity lower than Y1388 by about 0.05 dex ($\lesssim$ 2\,$\sigma$), although its abundance behaviour seems similar to that of Y1388. This star has RV of $\sim$ 20 km/s which is about 12 km/s lower than the average RV of other programme stars. This indicates that Y535 might has a wide binary companion, although it is difficult to estimate how this companion would affect the chemical composition of Y535.

Combining the fact that all three turn-off stars are high probability members of M67, our results imply that the turn-off stars in M67 are chemically inhomogeneous. This phenomenon has been reported for main-sequence dwarf stars in the Hyades \citep{liu16a} and in the Pleiades \citep{spi18}, although those results were explained using different scenarios. Further discussion about abundance trends versus dust condensation temperature for our results are presented in Appendix B.

\begin{table*}
\caption{The average abundances, related dispersions (standard deviations of the mean), and the corresponding average errors for the sub-giant stars and turn-off stars in our sample for 24 species, relative to the Sun.}
\centering
\label{t:stat}
\begin{tabular}{@{}cccc|ccc@{}}
\hline
 & \multicolumn{3}{c|}{Sub-giant stars} & \multicolumn{3}{c}{Turn-off stars} \\
Species & $<$[X/H]$>$ & $\sigma(<$[X/H]$>$)$^a$ & $<\sigma$[X/H]$>^b$ & $<$[X/H]$>$ & $\sigma(<$[X/H]$>$)$^a$ & $<\sigma$[X/H]$>^b$ \\
\hline
 C\,{\sc i}  & 0.049 & 0.039 & 0.018 & $-$0.054 & 0.063 & 0.023 \\
 O\,{\sc i}  & 0.046 & 0.023 & 0.025 & $-$0.040 & 0.060 & 0.023 \\
Na\,{\sc i}  & 0.066 & 0.013 & 0.016 & $-$0.062 & 0.074 & 0.012 \\
Mg\,{\sc i}  & 0.119 & 0.015 & 0.008 & $-$0.039 & 0.095 & 0.012 \\
Al\,{\sc i}  & 0.076 & 0.020 & 0.010 & $-$0.107 & 0.084 & 0.019 \\
Si\,{\sc i}  & 0.101 & 0.015 & 0.011 & $-$0.035 & 0.075 & 0.010 \\
 S\,{\sc i}  & 0.086 & 0.015 & 0.026 & $-$0.071 & 0.037 & 0.026 \\
Ca\,{\sc i}  & 0.083 & 0.010 & 0.014 & $-$0.030 & 0.073 & 0.017 \\
Sc\,{\sc ii} & 0.094 & 0.009 & 0.025 & $-$0.058 & 0.089 & 0.025 \\
Ti\,{\sc i}  & 0.056 & 0.007 & 0.023 & $-$0.057 & 0.063 & 0.022 \\
Ti\,{\sc ii} & 0.090 & 0.013 & 0.023 & $-$0.076 & 0.086 & 0.029 \\
 V\,{\sc i}  & 0.099 & 0.025 & 0.025 & $-$0.045 & 0.066 & 0.028 \\
Cr\,{\sc i}  & 0.061 & 0.014 & 0.019 & $-$0.067 & 0.064 & 0.024 \\
Mn\,{\sc i}  & 0.031 & 0.030 & 0.024 & $-$0.113 & 0.089 & 0.028 \\
Fe\,{\sc i}  & 0.079 & 0.006 & 0.013 & $-$0.050 & 0.087 & 0.015 \\
Fe\,{\sc ii} & 0.078 & 0.006 & 0.021 & $-$0.055 & 0.087 & 0.020 \\
Co\,{\sc i}  & 0.062 & 0.030 & 0.022 & $-$0.050 & 0.053 & 0.019 \\
Ni\,{\sc i}  & 0.077 & 0.004 & 0.014 & $-$0.074 & 0.087 & 0.018 \\
Cu\,{\sc i}  & 0.088 & 0.024 & 0.035 & $-$0.087 & 0.112 & 0.028 \\
Zn\,{\sc i}  & 0.007 & 0.024 & 0.024 & $-$0.133 & 0.079 & 0.024 \\
Sr\,{\sc i}  & 0.052 & 0.014 & 0.032 & $-$0.041 & 0.054 & 0.032 \\
 Y\,{\sc ii} & 0.080 & 0.019 & 0.028 & $-$0.080 & 0.096 & 0.025 \\
Ba\,{\sc ii} & 0.109 & 0.009 & 0.026 & $-$0.041 & 0.100 & 0.034 \\
Ce\,{\sc ii} & 0.081 & 0.005 & 0.036 & $-$0.095 & 0.063 & 0.035 \\
\hline
\end{tabular}
\\
$^a$ Abundance dispersions: the standard deviations of the mean abundances.\\
$^b$ Average errors: the mean of errors associated with a particular species.
\end{table*}

\subsection{Comparison to the previous studies of M67}

\begin{figure}
\centering
\includegraphics[width=\columnwidth]{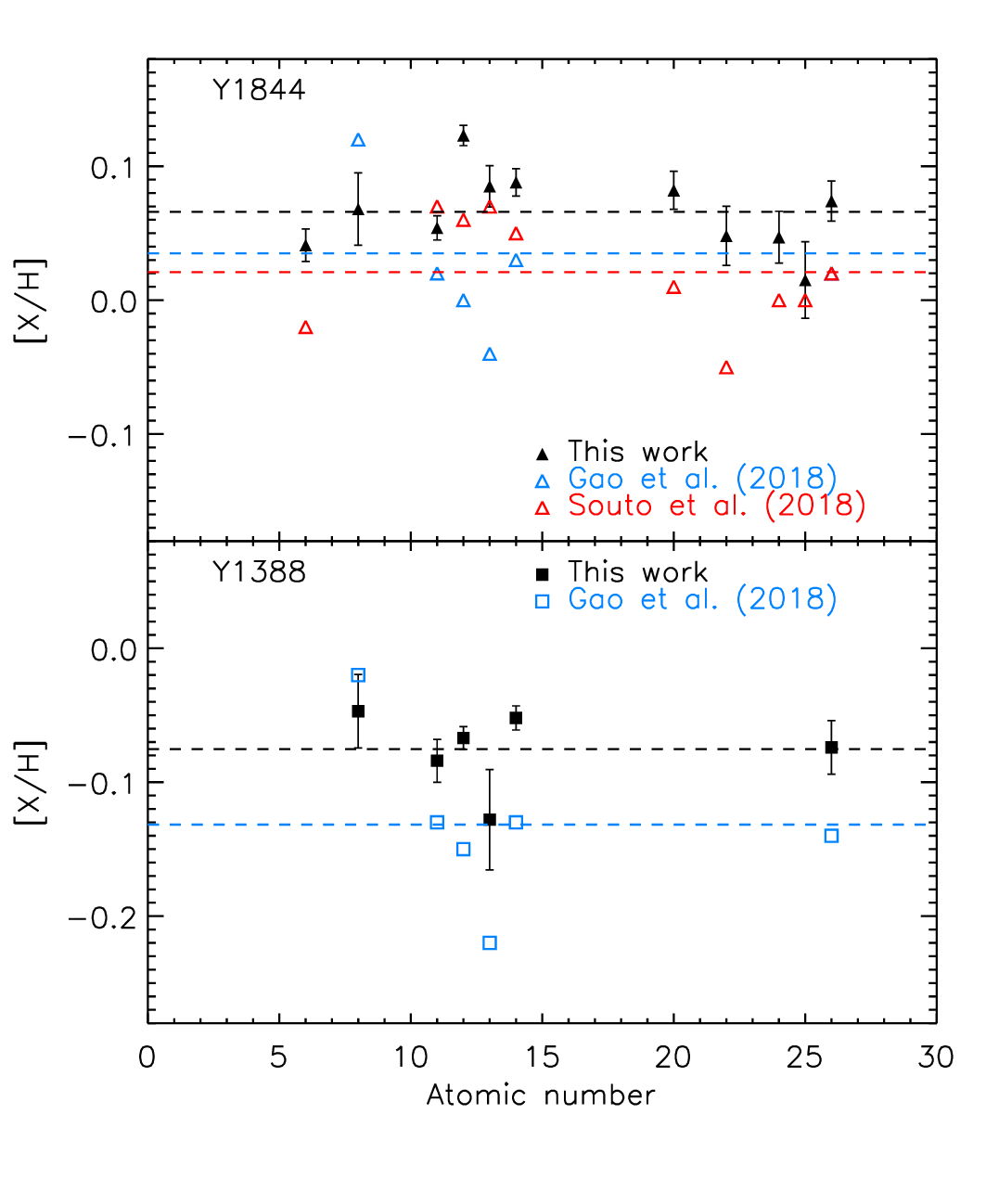}
\caption{Comparison of our results with those from \citet{gao18} and \citet{sou18} for the two stars in common. Top panel: Comparison of elemental abundances of the sub-giant star Y1844; black, blue, and red triangles represent the results from this work, \citet{gao18}, and \citet{sou18}, respectively. Bottom panel: Comparison of elemental abundances of the turn-off star Y1388; black and blue rectangles represent the results from this work and \citet{gao18}, respectively. The dashed lines mark the locations of mean elemental abundances from the corresponding studies.}
\label{fig7}
\end{figure}

Before discussing the implications of our results, we compare our results to the previous spectroscopic studies of M67. First, we compare our results to those from \citet{gao18}\footnote{We adopted their NLTE results for all the common elements.} for the two stars in common: Y1844 (sub-giant) and Y1388 (turn-off), and to that from \citet{sou18}\footnote{Only LTE results are available and thus being adopted.} for the one sub-giant stars we have in common: Y1844 (see Figure \ref{fig7}). For Y1844, our derived elemental abundances for elements in common are more metal-rich by: 0.057 $\pm$ 0.027 ($\sigma$ = 0.065) dex and 0.045 $\pm$ 0.010 ($\sigma$ = 0.033) dex, when compared to that from \citet{gao18} and \citet{sou18}, respectively. For Y1388, our elemental abundances for common elements are also more metal-rich than that from \citet{gao18} by: 0.056 $\pm$ 0.018 ($\sigma$ = 0.044) dex. For individual star, our results agree with these two studies within the errors, since the typical errors in the elemental abundances are $\sim$ 0.05 - 0.1 dex in these two studies we compare with. 

\begin{figure}
\centering
\includegraphics[width=\columnwidth]{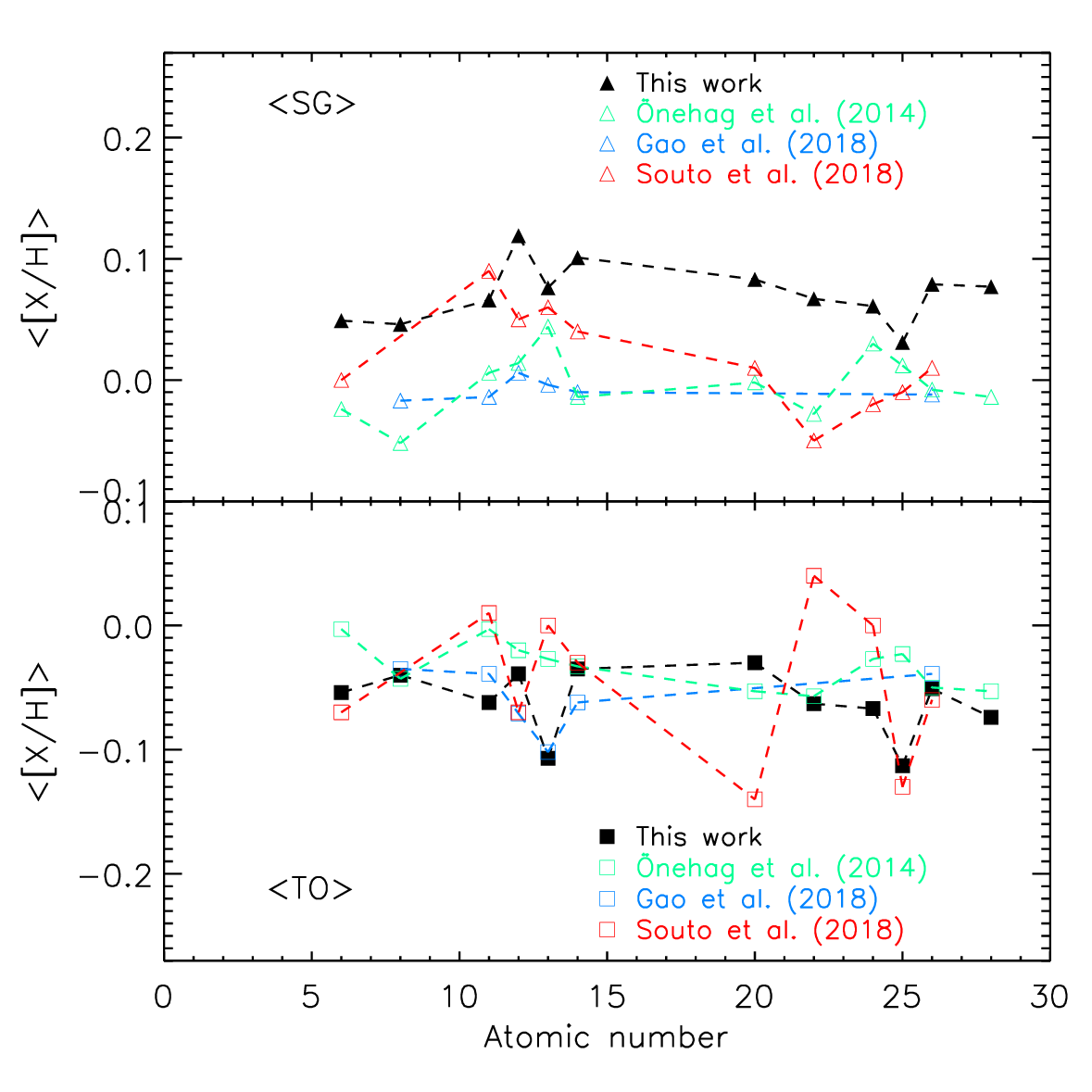}
\caption{Comparison of our results with those from \citet{one14}, \citet{gao18}, and \citet{sou18} for the average elemental abundances for sub-giant and turn-off stars. Top panel: Comparison of the average elemental abundances for the sub-giant stars; Bottom panel: Comparison of the average elemental abundances for the turn-off stars. Black, green, blue, and red symbols represent the results from this work, \citet{one14}, \citet{gao18}, and \citet{sou18}, respectively.}
\label{fig8}
\end{figure}

Secondly, we compared the average elemental abundances of our stars to those from \citet{one14}, \citet{gao18}, and \citet{sou18} for the sub-giant and turn-off stars separately (see Figure \ref{fig8}). We found that the average elemental abundances for the sub-giant stars in our study are higher by about 0.06 dex, 0.08 dex and 0.05 dex, when compared to that from \citet{gao18} and \citet{sou18}, respectively. Meanwhile the average elemental abundances for the turn-off stars in this study is slightly lower than that from \citet{one14} by about 0.03 dex, similar to that from \citet{gao18}, and lower than that from \citet{sou18} by about 0.02 dex. We note that the average difference in elemental abundances between the sub-giant and turn-off stars in our study is more significant by $\approx$ 0.07 - 0.08 dex, when compared to the two previous studies. Such a difference in elemental abundance between sub-giant and turn-off stars is marginally comparable when taking into account the measurement uncertainties (and abundance scatters). The difference mainly originates from the quality of spectra and the level of uncertainties, as well as the selection of stars. For example, compared to the sample of our sub-giant stars, the sub-giants selected in \citet{one14} have in general similar $\log g$ but hotter temperature by $\sim$ 400 K. Therefore their sub-giants are located on an earlier stage on the stellar evolutionary tracks, which partly explain why the abundance difference reported in their study is smaller.

In general, our results clearly revealed the existence of abundance differences between the sub-giant stars and the turn-off stars, where the sub-giant stars are more metal-rich than the turn-off stars. Although the observed abundance differences in this study (0.1 - 0.2 dex) are larger with higher significance than the previous spectroscopic studies. We recall that the typical uncertainty in abundance for individual star from our study is $\sim$ 0.02 dex, while the typical errors in abundances are between 0.05 - 0.15 dex in the previous studies (e.g., \citealp{one14,gao18,sou18,sou19,mot18}). 

\subsection{Effects of atomic diffusion on derived elemental abundances}
\label{sect:atomdiff}

As noted in the introduction, atomic diffusion can cause surface abundances to vary as a function of stellar evolutionary phase. The maximum effect is seen near the turn-off point. As the star evolves from the turn-off point to the sub-giant and red giant branches, the surface abundances return to (near) their original values due to the deepening of the surface convection zone.

From this study, we see that the sub-giant stars in our sample are chemically homogeneous, to a high-precision level ($\approx$ 0.02 dex). Since the convection zones of sub-giant stars are deep, any signatures of diffusion might be washed away. It is therefore also likely that the chemical composition of these stars is the closest to that of the primordial composition of the cluster based on this study. We note that the turn-off stars in our sample show large abundance variations (0.1 - 0.2 dex), implying that they are likely chemically inhomogeneous. Since the turn-off stars have the thinnest convective envelopes and diffusion has had the whole time of the main sequence evolution to affect their surface abundances, thus making their surface abundances variable. This might lead to the observed large abundance variations between the turn-off stars from our sample. We therefore suspect that the turn-off stars in M67 might show the largest chemical inhomogeneity, when compared to other stellar groups. 

\begin{figure*}
\centering
\includegraphics[width=\textwidth]{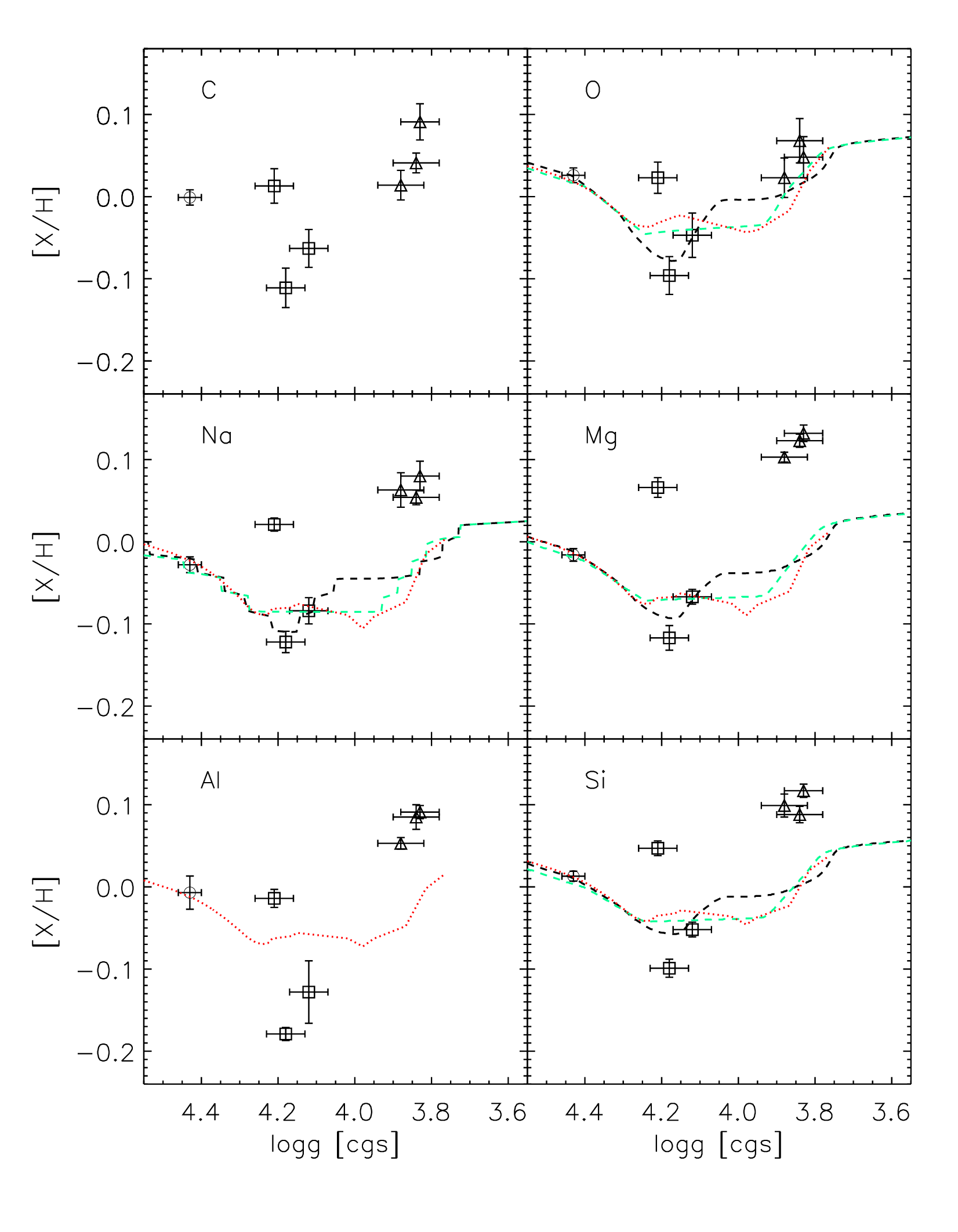}
\caption{Elemental abundances of C, O, Na, Mg, Al, and Si for our sample stars. The black and green dashed lines represent the model predictions with solar metallicity, and an age of 4.0 Gyr and 5.0 Gyr, respectively \citep{dot17}. The red dotted line represents the model predictions by O. Richard (priv. comm.). Open triangles and rectangles represent the elemental abundances for the sub-giant and turn-off stars, respectively. The solar twin Y1194 is shown as an open circle.}
\label{fig9}
\end{figure*}

\begin{figure*}
\centering
\includegraphics[width=\textwidth]{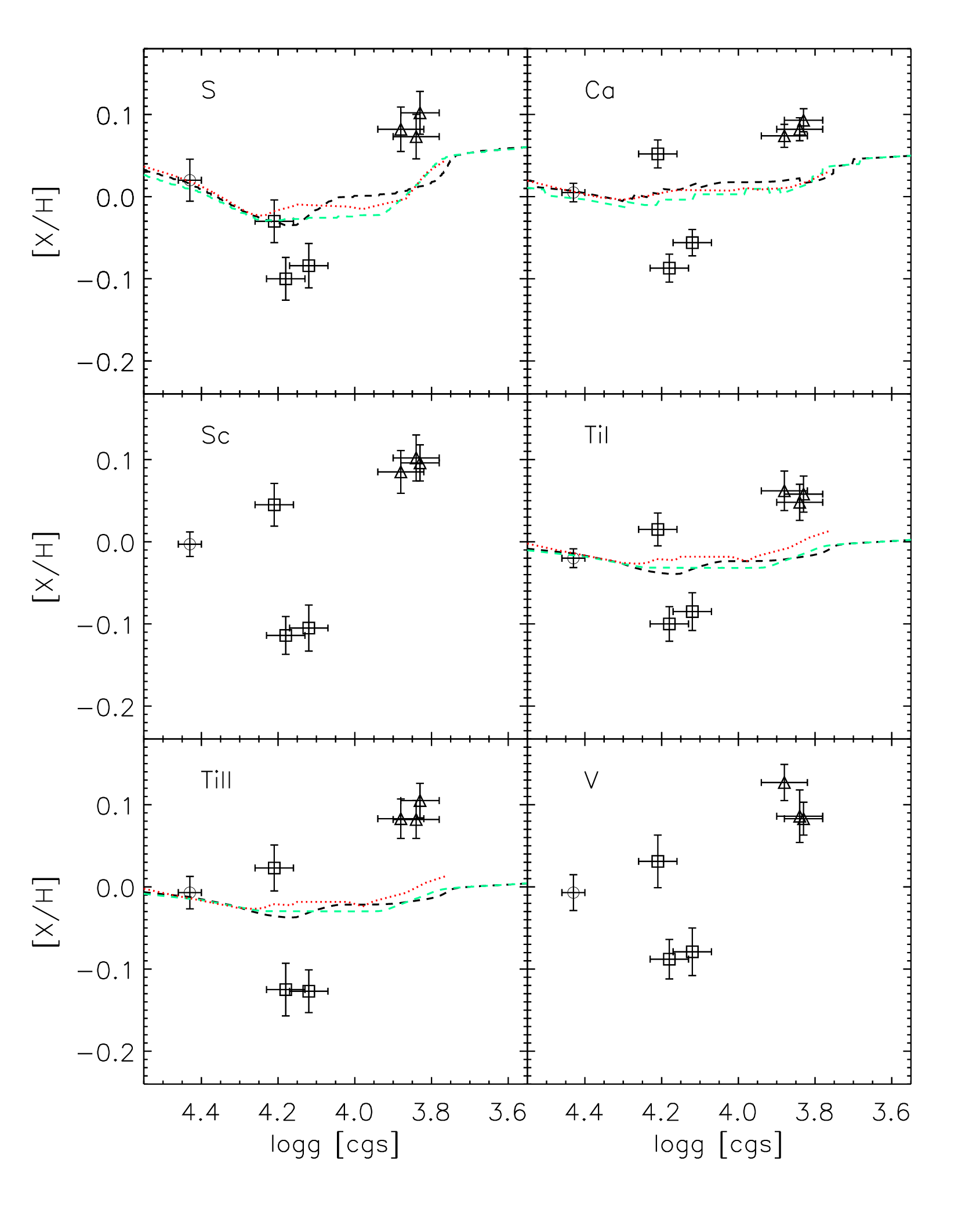}
\caption{Same as Figure \ref{fig9} but for S, Ca, Sc, Ti\,{\sc i}, Ti\,{\sc ii}, and V.}
\label{fig10}
\end{figure*}

\begin{figure*}
\centering
\includegraphics[width=\textwidth]{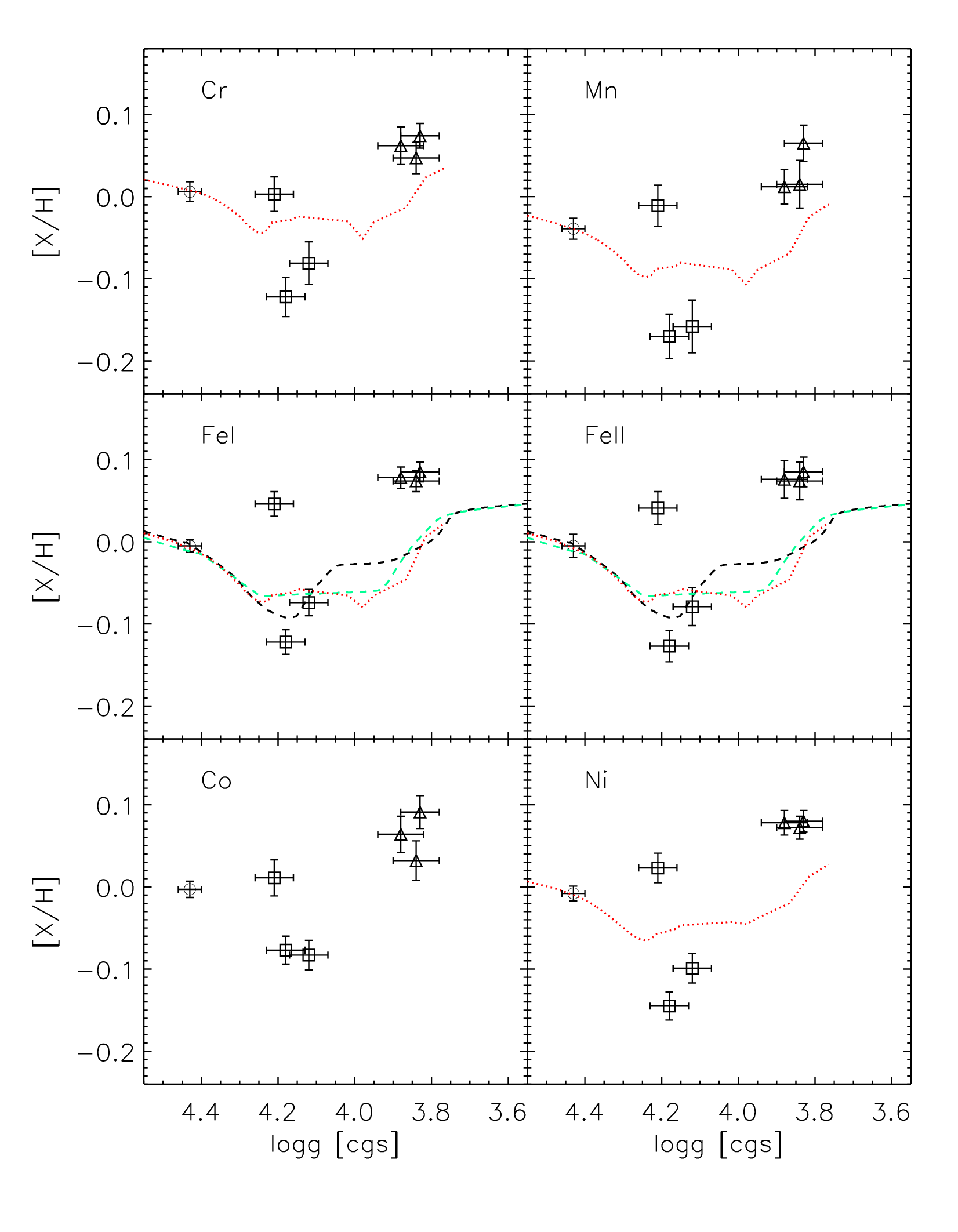}
\caption{Same as Figure \ref{fig9} but for Cr, Mn, Fe\,{\sc i}, Fe\,{\sc ii}, Co, and Ni.}
\label{fig11}
\end{figure*}

\begin{figure*}
\centering
\includegraphics[width=\textwidth]{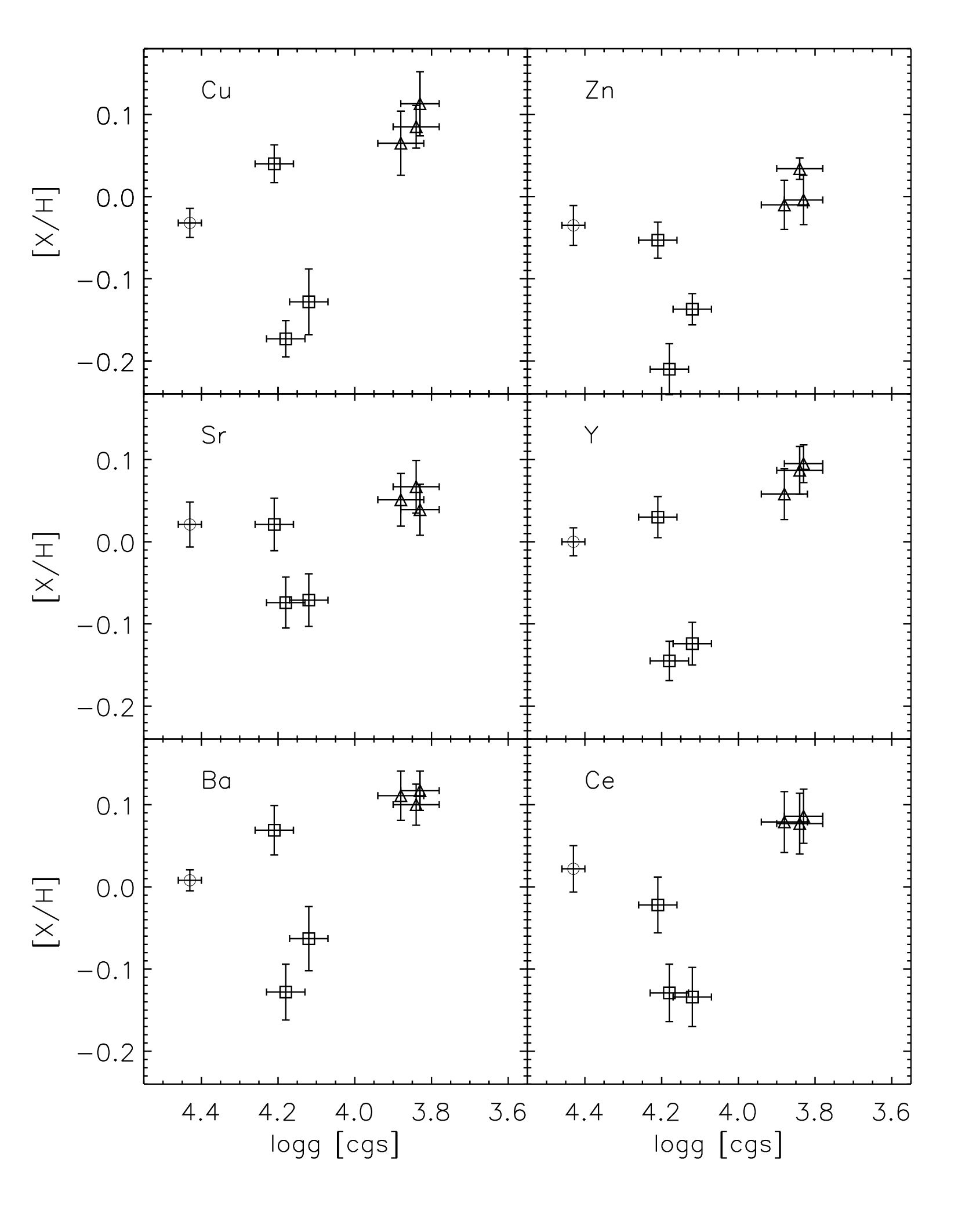}
\caption{Same as Figure \ref{fig9} but for Cu, Zn, Sr, Y, Ba, and Ce.}
\label{fig12}
\end{figure*}

In order to test the effect of atomic diffusion in M67, we compared our observational results to the stellar evolutionary models from \citet{dot17} and O. Richard (private communication). In Figures \ref{fig9} -- \ref{fig12}, we show the elemental abundances for all the species analysed in this study, as well as for the values of solar twin Y1194. We then over-plotted the values predicted by the model for those species which have been modelled. The adopted MIST models include overshooting mixing, turbulent diffusion and atomic diffusion, and radiative accelerations were included on an element-by-element basis. The elemental abundances predicted by the model were derived for solar metallicity \citep{asp09}, initial mass ranging from 0.7\,M$_\odot$ to 1.4\,M$_\odot$, and ages of t = 4.0 Gyr and t = 5.0 Gyr. The model from O. Richard has taken into account atomic diffusion, but without additional mixing. The predicted abundances by his model were based on an age of $\sim$ 3.7 Gyr and solar metallicity slightly differ ($\sim$ $+$0.06 dex) from the value adopted by MIST models. We shifted the zero-point of the model predicted values to fit the results for the solar twin Y1194. We note that there are subtle differences between different models but they agree in general.

For all elements, the relative elemental abundance ratios between the solar twin, the turn-off and sub-giant stars qualitatively matches the model. The turn-off stars have, on average, lower [X/H] than the solar twin while the sub-giant stars have, on average, higher [X/H] ratios than the solar twin. We note that our conclusion regarding diffusion does not hinge on a particular star such as the 'outlier' Y2235. The best agreement between model predictions and observations occurs for oxygen. For other elements, the observed abundance differences are in general larger than the model predictions. We note that radiative levitation and turbulent mixing were taken into account for the models. Both of them have the effect of reducing the diffusion signatures in the turn-off stars. It has generally been found that a small amount of surface mixing, namely "turbulent mixing", is required to match observations because the amount of depletion predicted by stellar models with diffusion but without any other mixing is too large. However, this extra mixing at the surface has not been well tested by any means and not well justified physically. A physically-motivated treatment of turbulent mixing in the surface layers would help in this regard. More observational results with high precision for the stars throughout the whole evolutionary stage are needed to better constrain/quantify the exact amount of turbulent mixing.

\subsection{The age and initial metallicity of M67}

As mentioned in the Introduction, M67 is an open cluster with near solar metallicity and an age of about 4 Gyr \citep{van07,yad08,sar09,one11,one14}. Numerous theoretical studies by e.g., \citet{van06,mag10,bre12,cho16} made use of it to calibrate the efficiency of convective overshoot mixing in low-mass stars. 
The age estimation of M67 could benefit from the high precision stellar spectroscopic parameters derived in a strictly differential approach. \citet{one11} estimated the age of a solar twin Y1194 in M67 to be: 4.2 $\pm$ 1.6 Gyr by fitting stellar evolutionary tracks using the Victoria stellar evolutionary code (based on \citealp{van07}) to their well determined stellar spectroscopic parameters.

\begin{figure}
\centering
\includegraphics[width=\columnwidth]{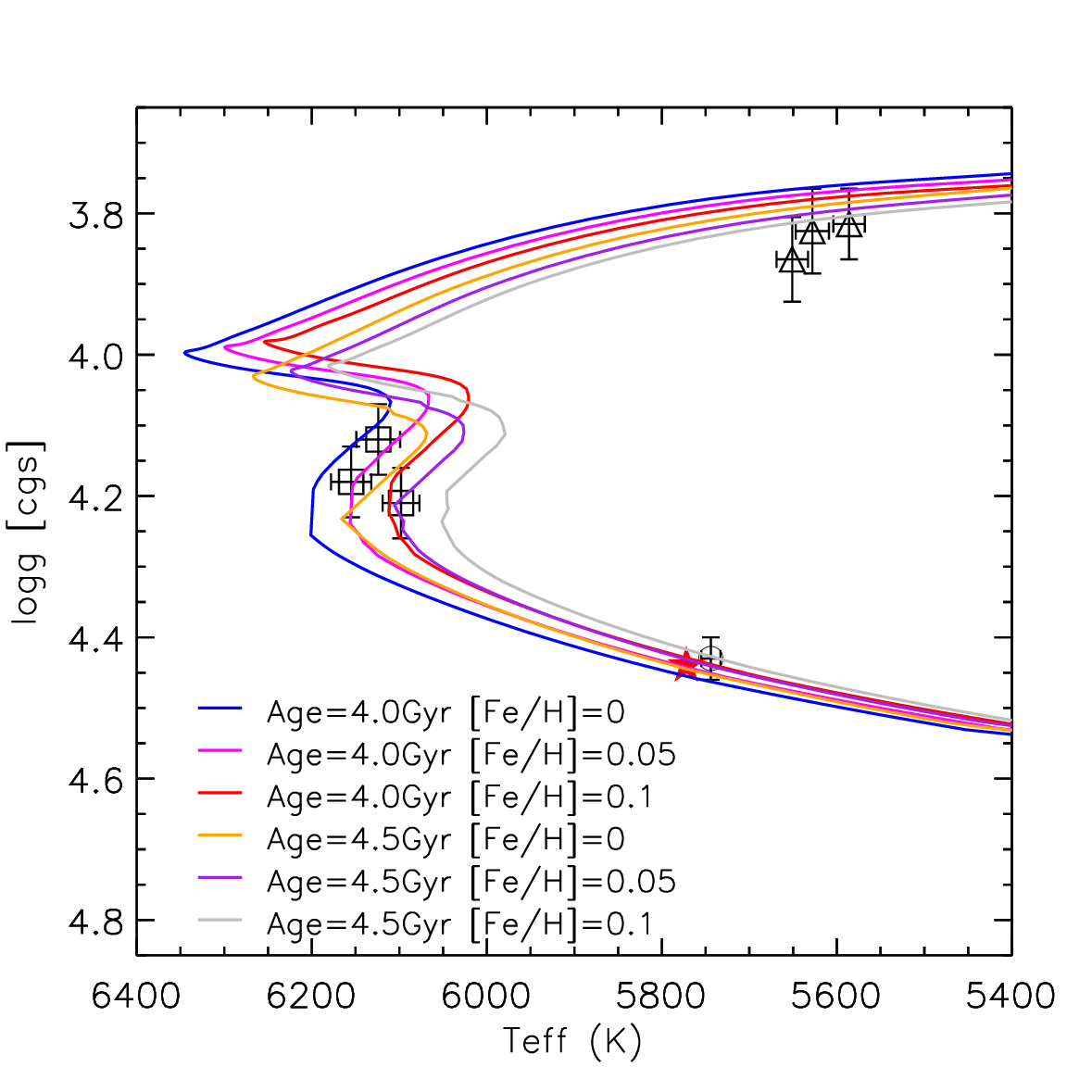}
\caption{Stellar spectroscopic parameters (T$_{\rm eff}$ and $\log g$) of our programme stars. The open circle, open rectangles, and open triangles correspond to our targets: solar twin Y1194, turn-off stars, and sub-giant stars, respectively. The solar value is marked as a blue star sign. A set of MIST isochrones with initial metallicities of [Fe/H] between 0 to 0.1, and age of 4 Gyr and 4.5 Gyr \citep{dot16,cho16} are shown as the solid lines.}
\label{fig13}
\end{figure}

In this study, we tried to fit the isochrones to our spectroscopic parameters for further exploration and verification. Figure \ref{fig13} shows our results compared to a set of MIST isochrones with initial metallicities of [Fe/H] between 0 to 0.1, and age of 4 Gyr and 4.5 Gyr \citep{dot16,cho16}. The values of spectroscopic $\log g$ of our sample stars seem slightly higher than the values from MIST models by $\sim$ 0.05 dex but they are still marginally comparable considering the uncertainties from observations as well as from theoretical models. It is difficult to tell which isochrone provides us the best fit. We would argue that, for the isochrones with an age of 4 Gyr, that [Fe/H] = $+$0.05 shows a good match for the two hottest stars while the other stars clearly lie nearer to [Fe/H] = 0.1; while for the isochrones with an age of 4.5 Gyr, that [Fe/H] = 0 shows a better match than [Fe/H] = 0.05 and 0.1. It has been suggested by \citet{one14} that the diffusion corrected initial metallicity of M67 is estimated to be [Fe/H] = 0.06. In this study, the average [Fe/H] of the sub-giant stars in M67 is $\sim$ 0.07 dex. Based on our observational results, in combination with the comparison with the theoretical isochrones, we estimate that the initial chemical composition of M67 is probably between 0.05 - 0.1. Nevertheless, \citet{cho18} discussed the theoretical uncertainties in the model-based T$_{\rm eff}$ scale. It is entirely possible that our T$_{\rm eff}$ scale and that of the models differ by some small amount ($\sim$ 25 - 50K) and this could cause a significant difference in the conclusion of which initial metallicity is to be preferred. In addition, the difference between different theoretical isochrones (e.g., MIST, PARSEC, and YY) would also affect the conclusion of the initial chemical composition of M67. 

Another method to estimate the stellar ages is using [Y/Mg] abundance ratios. \citet{nis15} and \citet{nis16} found that [Y/Mg] ratios correlate strongly with stellar ages for the solar twins and such abundance ratios can be used as a ’chemical clock’. The empirical relation between [Y/Mg] and age was presented in these two studies. \citet{fel17} confirmed the relation for dwarfs of solar metallicity but found that it disappears for stars with [Fe/H] $\sim$ $-$0.5 dex and below. \citet{slu17} claimed that the empirical relation between [Y/Mg] and age as presented by \citet{nis16} was found to hold also for helium-core-burning giants of close to solar metallicity.

In this study we derived the chemical age of a solar twin Y1194 in M67, using the [Y/Mg] ratio and the relation proposed by \citet{nis16}. We found the chemical age of Y1194 to be: 4.15 $\pm$ 0.59 Gyr. This agrees well with the age estimation using isochrone fitting to our results as well as that from \citet{one11}. However, for the turn-off and sub-giant stars in our sample, the chemical ages were found to be about 1.5 Gyr higher than that of the solar twin Y1194. The [Y/Mg] ratio in average is $\sim$ $-$0.03 dex for our sample stars, agree well with that from \citet{one14} ($\sim$ $-$0.04 dex). This is however lower than the value for a giant star in M67 (see Figure 3, \citealp{slu17}). Therefore we suspect that the empirical relation between [Y/Mg] and age might not work well for the turn-off and sub-giant stars.

\subsection{Implications for chemical tagging}

A basic assumption for chemical tagging to work is that open clusters, which are surviving star-forming aggregates in the Galactic disc, should be chemically homogeneous \citep{bla10,bc15,ds15,tin15}. Determining the level of chemical homogeneity in open clusters is thus of fundamental importance to study the evolution of star-forming clouds and that of the Galactic disc \citep{bov16}.

In the current study, we have focused on the type of stars that are the prime targets for chemical tagging -- dwarf stars on the main sequence, around the turn-off and stars on the sub-giant branch. Ideally, stars in these different evolutionary phases should, if they formed from a single chemically homogenous star forming event, have the same elemental abundances. Our target stars are all high-probability members of the old open cluster M67. 

We find that basically all elemental abundances in M67 vary as a function of evolutionary stage of the star. This is shown in Figures. \ref{fig9} -- \ref{fig12}, where we show the elemental abundances as a function of the surface gravity of the stars. The main sequence star in the sample has essentially solar elemental abundances, stars at the turn-off dip abut 0.1\,dex below solar values (but note the one star that differs, see discussion in Sect\,\ref{sect:atomdiff}), while the sub-giant stars again rise, now to super-solar values. Our study is of a highly differential nature, reaching high-precision in the elemental abundances. Such precision is rarely, if at all, reached in large spectroscopic surveys such as the Gaia-ESO survey \citep{gil12}, the GALAH survey \citep{ds15}, and the APOGEE survey \citep{maj17}, due to the lower quality spectra (relative to our study).

From this study, we found that the sub-giant stars in our sample are chemically homogeneous, under a high-precision level ($\approx$ 0.02 dex). As discussed in Sect\,\ref{sect:atomdiff} we expect stars in this evolutionary phase to have well-mixed atmospheres and they are thus likely to have chemical compositions close to the primordial composition of the cluster ([Fe/H] $\sim$ 0.05 - 0.1). This makes sub-giant stars ideal targets for the application of the chemical tagging technique. We find that our turn-off stars show large abundance variations (0.1 - 0.2 dex). Again, as discussed in Sect\,\ref{sect:atomdiff} this is likely because diffusion has had the whole time of the main sequence evolution to act on their surface abundances, as well as their thin convective envelopes, leading to the large variations in elemental abundances inside this group of stars. If this is true, then turn-off stars are probably not ideal targets for the chemical tagging technique and the purpose of Galactic archeology, unless the effects of atomic diffusion can be accurately corrected.
 
We note that this work is limited by the small number of stars studied. A high precision differential abundance analysis on more M67 members (especially turn-off stars), is necessary to probe and test the possible chemical inhomogeneity in the M67 turn-off stars. Still, we believe that some general comments on which stars to choose for chemical tagging may already be done now (keeping in mind that we have not yet probed the chemical homogeneity of the main-sequence stars). First, it would appear that sticking to one type of star might be highly advantageous if we want to find stars formed from the same star-formation event and that the best type of star appears to be the sub-giant stars in our study. This is a somewhat sub-optimum conclusion as sub-giant stars are, due to the short duration of this stellar evolutionary phase, relatively rare stars (e.g., see Figure \ref{fig1}). Technically, it should be possible to correct the elemental abundances for the turn-off stars to reflect initial composition of the star-forming event, however, given the yet limited number of models and knowledge about how the exact mixture of different elements in the atmosphere influences processes like diffusion such corrections would at best have rather poor precision, which would mean that the corrected elemental abundances are not good enough for chemical tagging to work. 

\section{Conclusions}

We presented a strictly line-by-line differential chemical abundance analysis of three turn-off stars and three sub-giants in M67, in order to confirm and quantify more precisely the effect of atomic diffusion in this old benchmark open cluster. Our targets are all high probability members of M67, according to \citet{yad08} and \citet{gai18}. We obtained high precision ($\sim$ 0.02 dex) relative abundance measurements for 22 elements for the sample stars.

We found firstly, that the sub-giant stars in our sample are chemically homogenous at a level of 0.02 dex, with an average elemental abundance of 0.074 dex, with an abundance dispersion of 0.016 dex and an average error in the elemental abundances of 0.022 dex. These stars are mirrors reflecting closely the original chemical composition of the open cluster, making them ideal targets for the application of chemical tagging. Secondly, we note that the turn-off stars in our sample show significant variations in the elemental abundances, with an average elemental abundance of $-$0.065 dex, with an average dispersion of 0.076 dex and an average error in elemental abundance of 0.023 dex. Our results imply that the turn-off stars in our sample are likely chemically inhomogeneous, possibly due to the influence of turbulent mixing and diffusion in their interiors. If confirmed by future studies, this poses a major challenge to the concept of chemical tagging, when using the turn-off as tracers or a mixture of stars from different evolutionary stages. Finally, we clearly showed that the sub-giant stars are overall more metal-rich than the turn-off stars by more than 0.1 dex, larger than the findings from previous spectroscopic studies. The relative elemental abundances for the solar twin, turn-off and sub-giant stars are qualitatively in agreement with stellar models that include atomic diffusion. We also find that the amplitude of the differences between the turn-off and sub-giant stars is larger than the model predictions. Our results indicate that the effect of atomic diffusion, resulting in changes in surface abundances of stars during the different evolutionary phases, is prominent and should be taken into account when applying the chemical tagging technique to track the stars at different evolutionary stage.

Again, we stress that the results reported here are based on a small number of stars. The most urgent action is now to extend the sample of stars studied in this way in M67 as well as in a few selected open clusters with somewhat different properties (higher and lower iron abundance and lower ages) to further strengthen the empirical evidence for the size of atomic diffusion on final derived elemental abundance and to probe the level of chemical (in)homogeneity in the open cluster. 

\section*{Acknowledgments}

F.L. and S.F. acknowledge support by the grant "The New Milky Way" from the Knut and Alice Wallenberg Foundation and the grant 184/14 from the Swedish National Space Agency. F.L. was also supported by the M\"arta and Eric Holmberg Endowment from the Royal Physiographic Society of Lund. This work has been supported by the Australian Research Council (grants FL110100012, FT140100554 and DP120100991). Parts of this research were conducted by the Australian Research Council Centre of Excellence for All Sky Astrophysics in 3 Dimensions (ASTRO 3D), through project number CE170100013. J.M. acknowledges support by FAPESP (2018/04055-8).
We acknowledge valuable discussions with Dr. Ross Church. We thank the referee for the suggestions/comments that helped us to improve our manuscript.
The authors thank the ANU Time Allocation Committee for awarding observation time to this project. The authors wish to acknowledge the very significant cultural role and reverence that the summit of Mauna Kea has always had within the indigenous Hawaiian community. We are most fortunate to have the opportunity to conduct observations from this mountain.

\begin{appendix}

\section{Supplementary material}

The following supplementary material is available for this article online:

Table A1. Atomic line data, as well as the measured equivalent widths, adopted for our analysis.

\section{Elemental abundances versus T$_{\rm cond}$}

\citet{mel09} conducted a high precision differential abundance analysis for the Sun and the solar twins. They found that the Sun is depleted in refractory elements (with high T$_{\rm cond}$) when compared to the majority ($\approx$ 85\%) of solar twins. The observed abundance differences correlate with T$_{\rm cond}$, that were attributed to the influence of terrestrial planet formation. With a large sample, \citet{bed18} confirmed that the Sun indeed has a special abundance pattern, when compared to the other Sun-like stars. However, the hypothesis proposed by \citet{mel09} for the solar abundance pattern is still under debate. \citet{adi14} and \citet{nis15} claimed that the detected abundances versus T$_{\rm cond}$ trends might be due to the effect of Galactic chemical evolution. Another possible scenario, namely 'dust-cleansing', suggested that the Sun was born in a dense environment where some of the dust in the proto-solar nebula was radiatively cleansed by luminous hot stars before the formation of the Sun, left its refractory elemental abundances deficient. This was proposed and discussed by \citet{one11} and \citet{one14} for their studies on M67 stars, where they found the M67 stars have chemical composition closer to the Solar composition than most solar twins. Their results indicate that the star forming location and environment might play an important role in explaining the solar abundance pattern.

\begin{figure}
\centering
\includegraphics[width=\columnwidth]{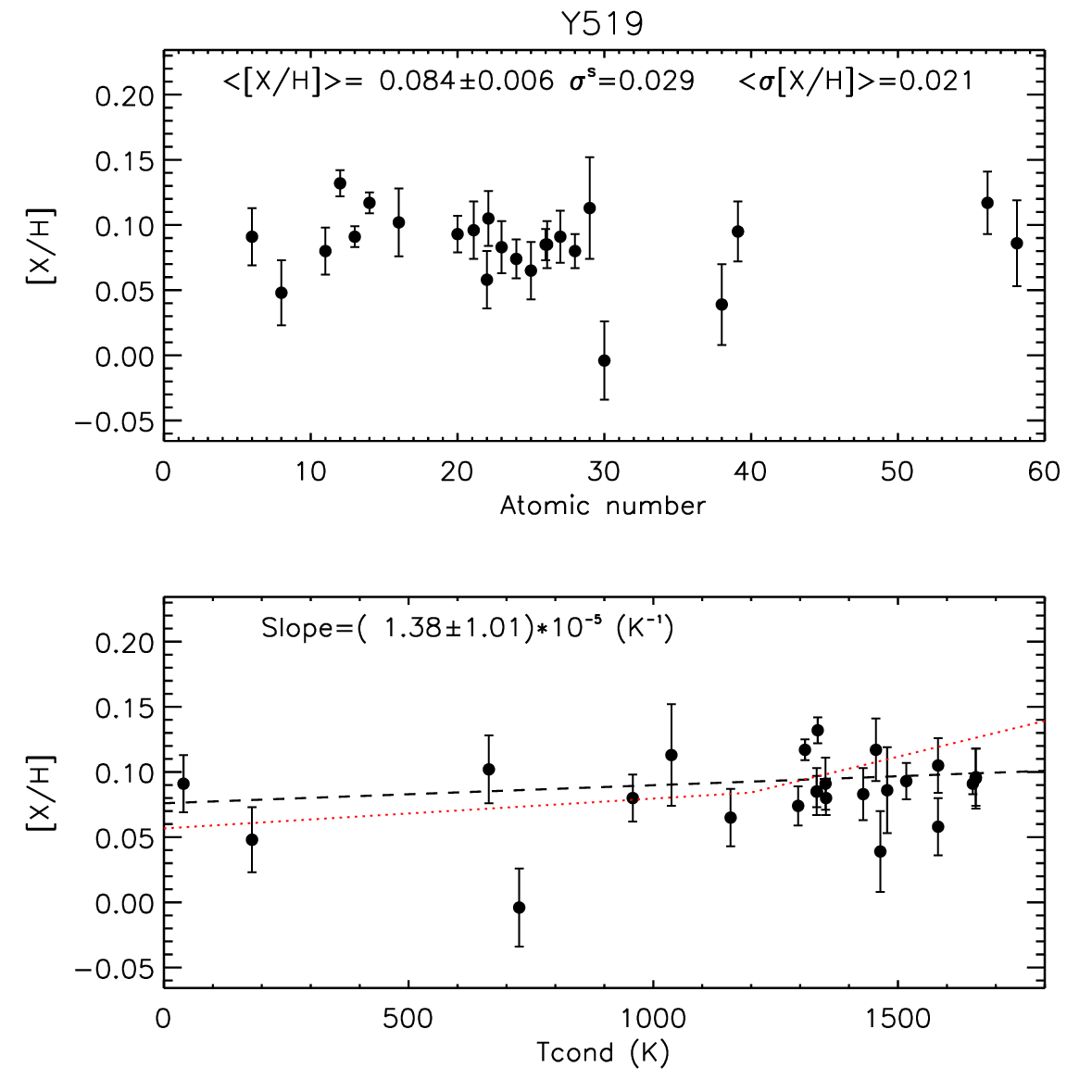}
\caption{Top panel: Differential abundances of a sub-giant star, Y519 relative to the Sun, as a function of atomic number. Bottom panel: Differential abundances of Y519 relative to the Sun, as a function of T$_{\rm cond}$; the blue dashed line represents the linear least-squares fit to the data and the red dotted line represents the T$_{\rm cond}$ trend taken from \citet{mel09}.}
\label{figb1}
\end{figure}

\begin{figure}
\centering
\includegraphics[width=\columnwidth]{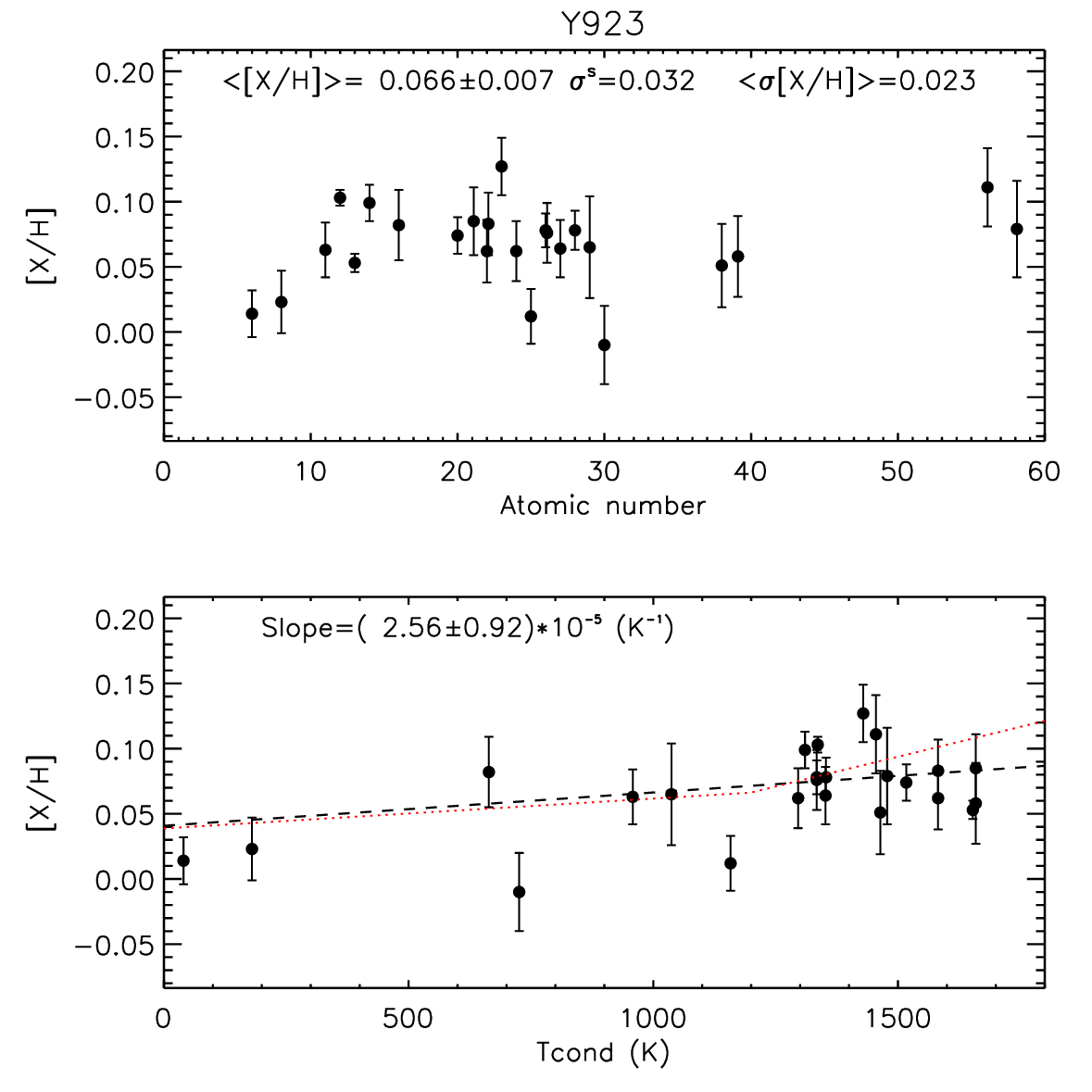}
\caption{Same as Figure \ref{figb1} but for a sub-giant star, Y923 relative to the Sun.}
\label{figb2}
\end{figure}

\begin{figure}
\centering
\includegraphics[width=\columnwidth]{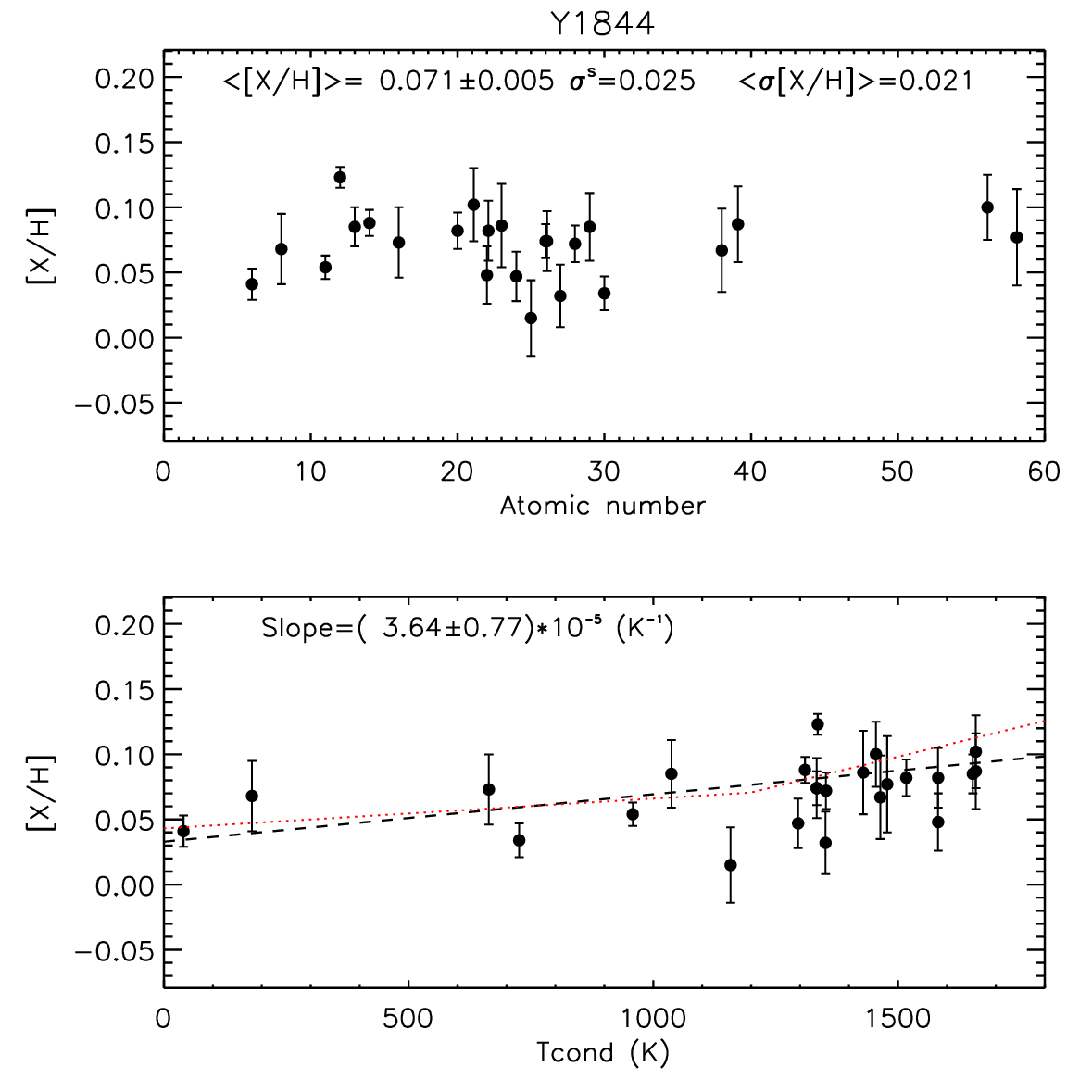}
\caption{Same as Figure \ref{figb1} but for a sub-giant star, Y1844 relative to the Sun.}
\label{figb3}
\end{figure}

\begin{figure}
\centering
\includegraphics[width=\columnwidth]{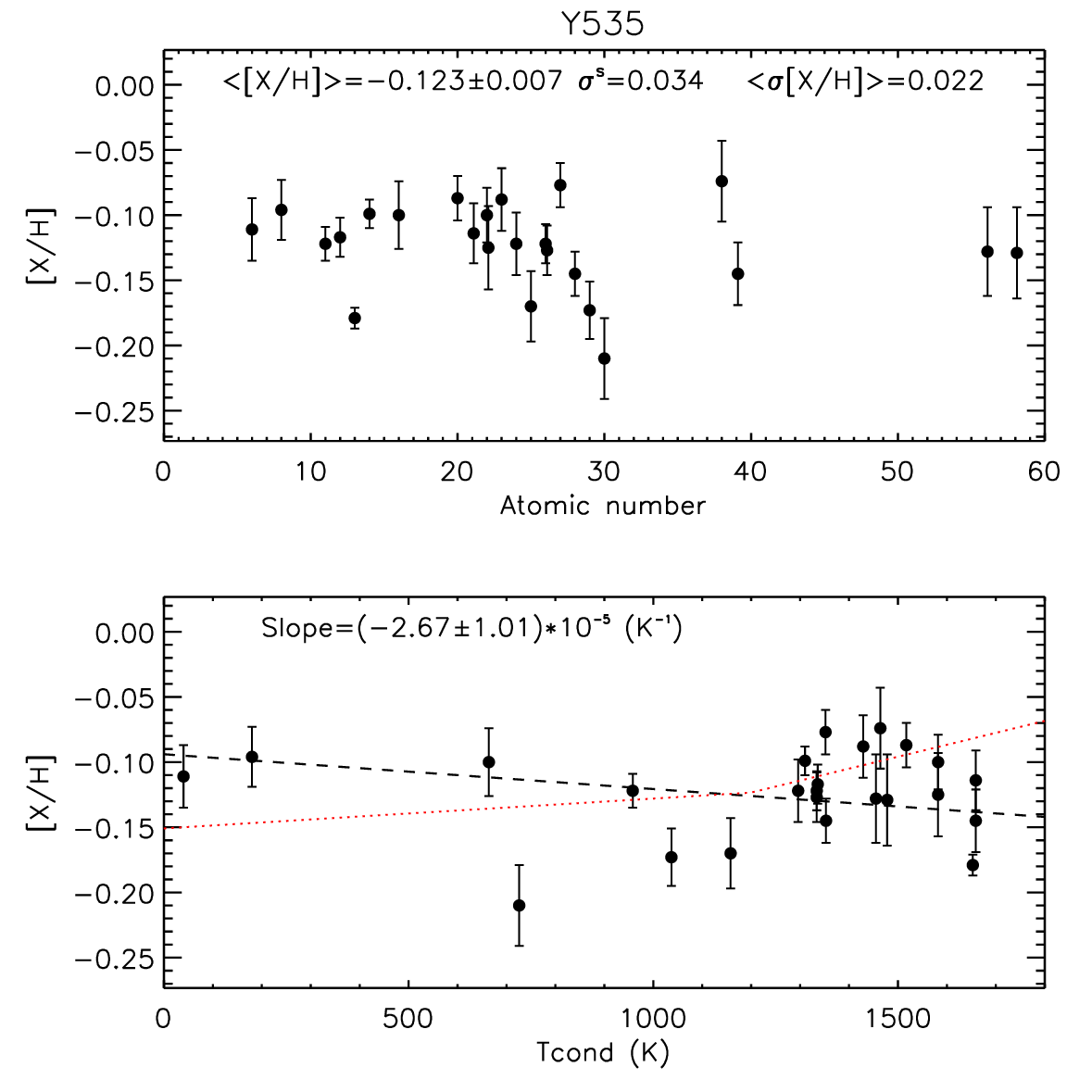}
\caption{Same as Figure \ref{figb1} but for a turn-off star, Y535 relative to the Sun.}
\label{figb4}
\end{figure}

\begin{figure}
\centering
\includegraphics[width=\columnwidth]{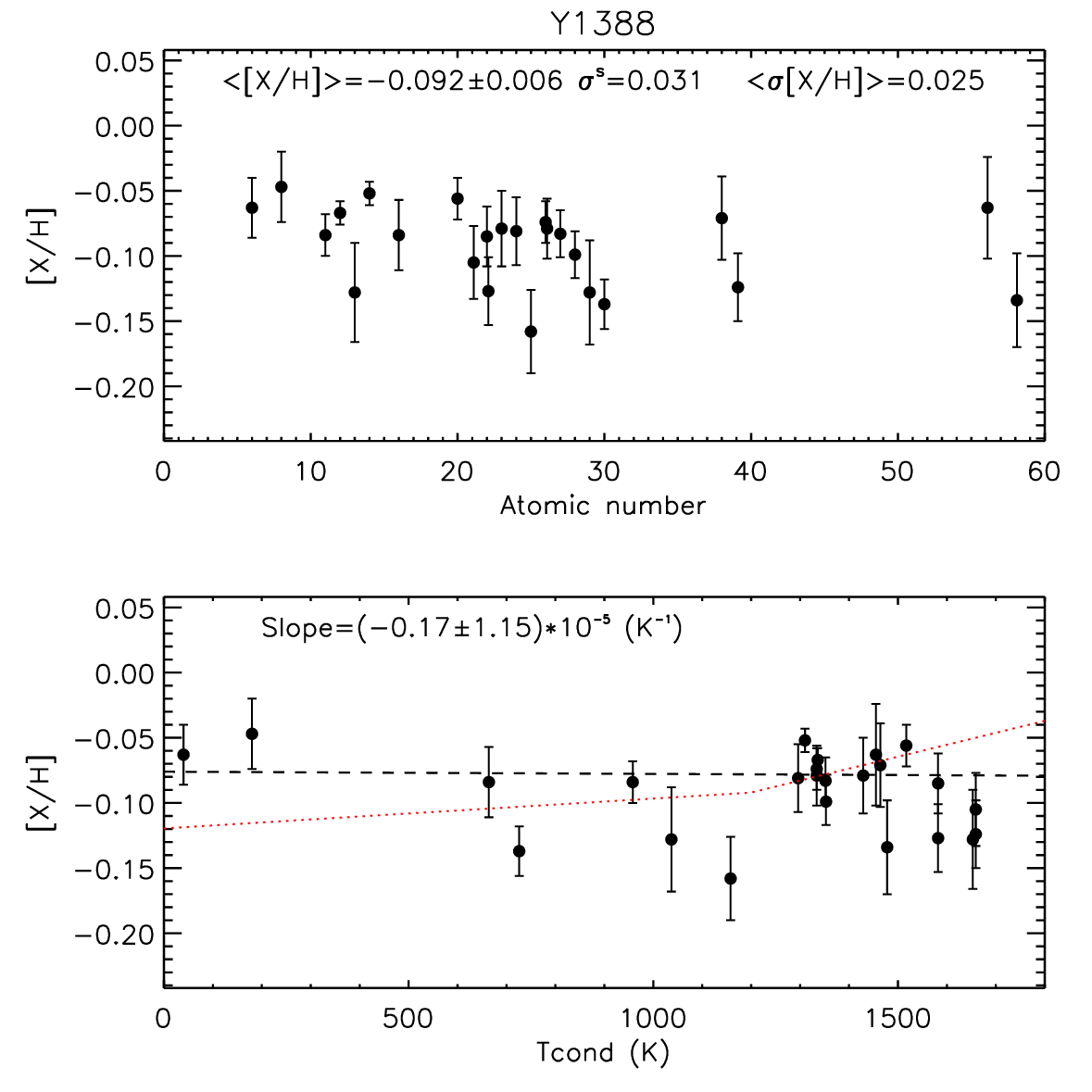}
\caption{Same as Figure \ref{figb1} but for a turn-off star, Y1388 relative to the Sun.}
\label{figb5}
\end{figure}

\begin{figure}
\centering
\includegraphics[width=\columnwidth]{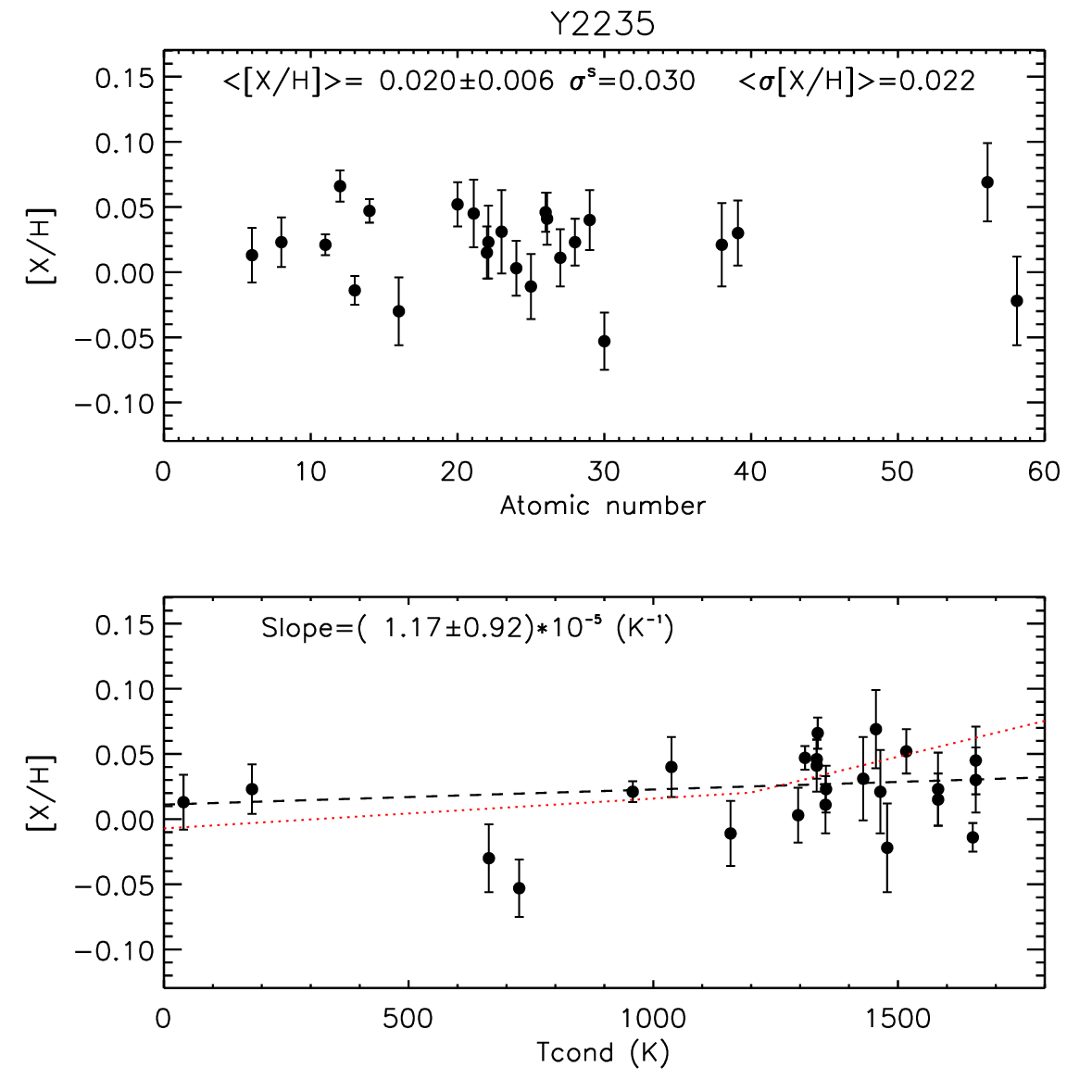}
\caption{Same as Figure \ref{figb1} but for a turn-off star, Y2235 relative to the Sun.}
\label{figb6}
\end{figure}

It is therefore of interest to examine the abundances versus T$_{\rm cond}$ trends for our sample stars. In Figures \ref{figb1}, \ref{figb2}, \ref{figb3}, \ref{figb4}, \ref{figb5}, and \ref{figb6} we show the differential elemental abundances relative to the Sun, as a function of T$_{\rm cond}$\footnote{50 \% T$_{\rm cond}$ of each element was taken from \citet{lod03}.}. We found that the slopes of [X/H] versus T$_{\rm cond}$ are positive with $>$ 2\,$\sigma$ significance level for two sub-giant stars (Y923 and Y1844), while relative flat for one sub-giant star (Y519). The slopes of [X/H] versus T$_{\rm cond}$ for two turn-off stars (Y1388 and Y2235) are flat, while negative with about 2.5\,$\sigma$ significance level for one turn-off star (Y535). Because of the relative large scatters in the [X/H] versus T$_{\rm cond}$ plots, we can not identify any chemical signatures due to planet formation. We note that the abundance behaviours of our sample stars agree well with that from \citet{one14}, slightly favour the 'dust-cleansing' hypothesis proposed in their study.

\begin{figure}
\centering
\includegraphics[width=\columnwidth]{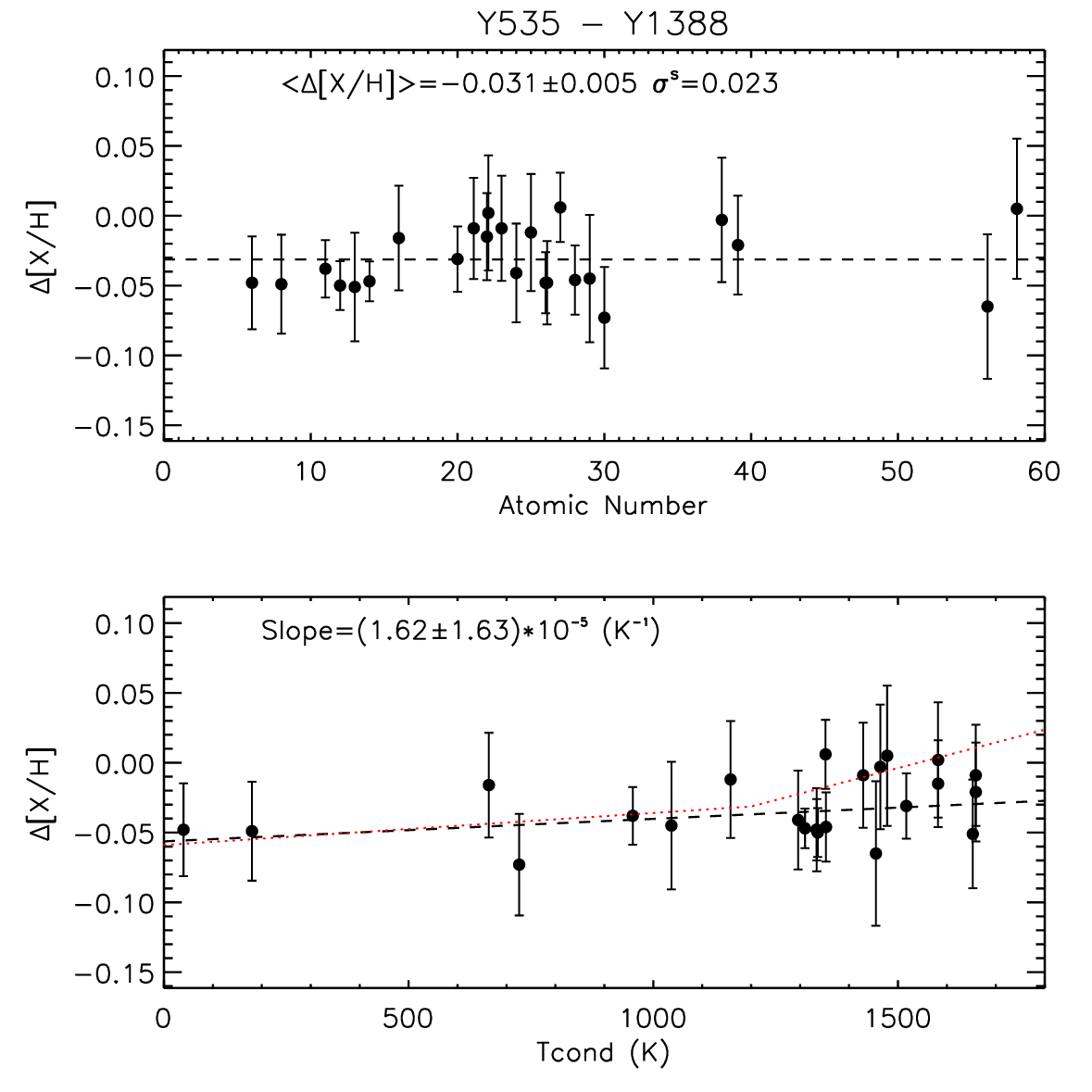}
\caption{Same as Figure \ref{figb1} but for Y535 relative to Y1388.}
\label{figb7}
\end{figure}

\begin{figure}
\centering
\includegraphics[width=\columnwidth]{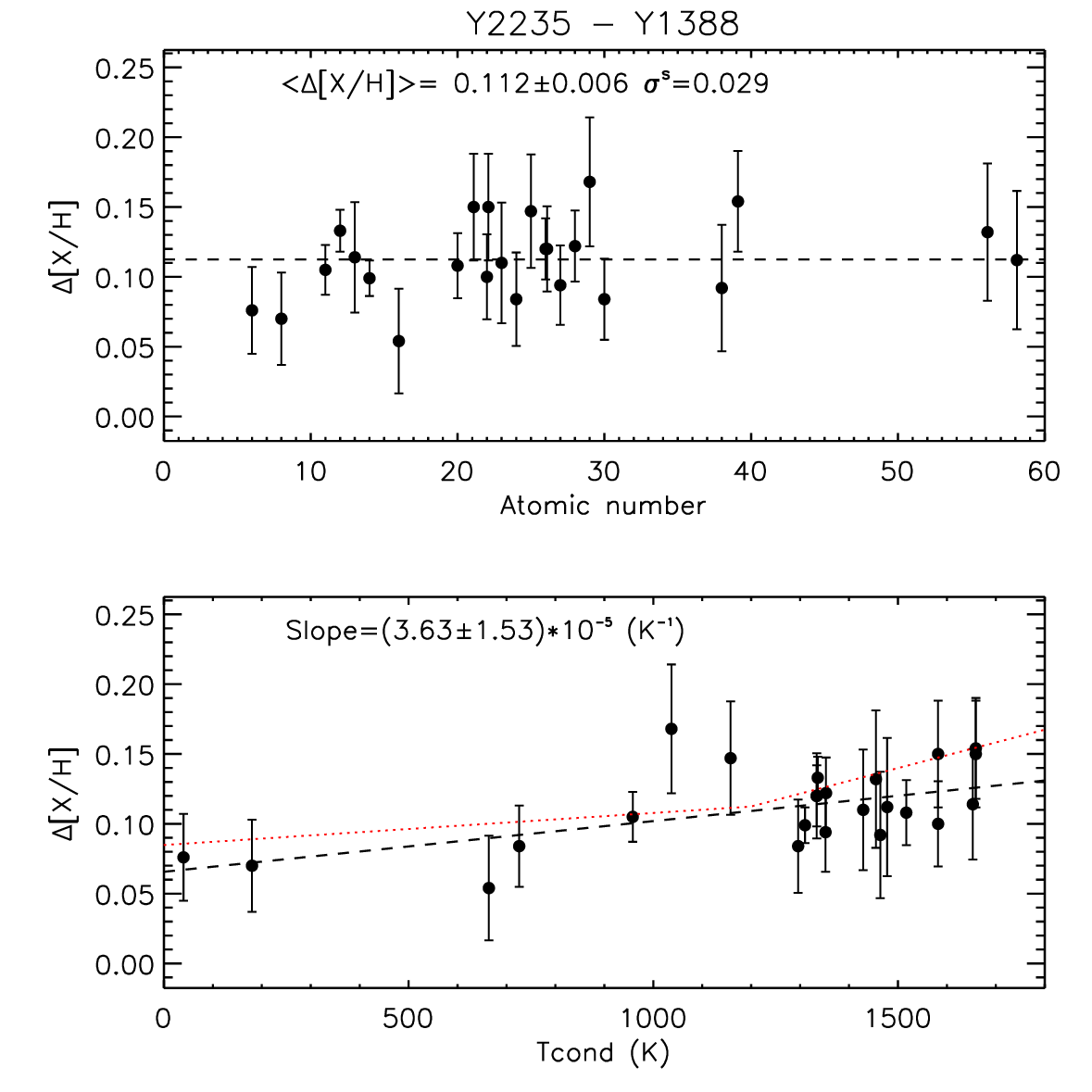}
\caption{Same as Figure \ref{figb1} but for Y2235 relative to Y1388.}
\label{figb8}
\end{figure}

Considering the fact that our turn-off stars are likely not chemically homogeneous, we examined the detailed abundance differences between these three turn-off stars. We show the differential elemental abundances of Y2235 and Y535 relative to Y1388 (the turn-off star with medium metallicity), as a function of T$_{\rm cond}$ in Figures \ref{figb7} and \ref{figb8}. We found that the slope of $\Delta$[X/H] versus T$_{\rm cond}$ is almost flat when compared Y535 to Y1388, while the corresponding slope is positive with 2.4\,$\sigma$ significance level when compare Y2235 to Y1388. The phenomenon are similar to the findings by \citet{spi18}, who found trends with condensation temperature in the differential abundances of Pleiades' stars, and suggested a possible connection with planets.

\end{appendix}

\end{document}